\documentclass[aip, pop, floats, superscriptaddress, twocolumn, floatfix]{revtex4}

\usepackage{graphicx}
\usepackage{bm}
\usepackage{amsfonts}
\usepackage[usenames, dvipsnames]{color}

\usepackage{amsmath}    
\usepackage{epsfig}
\usepackage{subfigure}  
\usepackage{hyperref}   
\usepackage{bm}
\usepackage{amssymb}

\newcommand{\p}[1]{\partial_{#1}}
\newcommand{\avg}[1]{\left\langle #1 \right\rangle}
\newcommand{\ou}{\overline{u}}
\newcommand{\ob}{\overline{b}}
\newcommand{\tij}[1]{\tau_{ij}^{#1}}
\renewcommand{\vec}[1]{\bm {#1}}
\newcommand{\fvec}[1]{\hat{\bm {#1}}}
\newcommand{\bu}{\vec{u}}
\newcommand{\bb}{\vec{b}}

\newcommand{\Pm}{\rm Pm}
\renewcommand{\Re}{{\rm Re}}
\newcommand{\Rl}{{\rm Re}_\lambda}

\newcommand{\blue}[1]{{\textcolor{black}{#1}}}

\newcommand{\ml}[1]{{\textcolor{black}{#1}}}
\newcommand{\lb}[1]{{\textcolor{black}{#1}}}
\newcommand{\mb}[1]{{\textcolor{black}{#1}}}

\newcommand{\marburgaddress}{Fachbereich Physik, Philipps-Universit\"at Marburg, Renthof 6, 35032 Marburg, Germany}

\newcommand{\romeaddress}{Department of Physics and INFN, University of 
Rome ``Tor Vergata'', Via della Ricerca Scientifica 1, 00133, Rome, Italy.}  

\newcommand{\tueaddress}{Department of Applied Physics, Eindhoven University of Technology, P.O. Box 513, 5600 MB Eindhoven, The Netherlands}

\begin{document}

\vspace{-3em}
\begin{minipage}[l]{\textwidth}
\noindent
Postprint version of the manuscript published in Physics of Plasmas
{\bf 25}, 122307 (2018). \\
\end{minipage}
\vspace{1em}

\title{{\em A-priori} study of the subgrid energy transfers for small-scale dynamo
in  kinematic  and saturation regimes}

\author{Gerardus Petrus Offermans}
\affiliation{\tueaddress}
\author{Luca Biferale} 
\affiliation{\romeaddress}
\author{Michele Buzzicotti} 
\affiliation{\romeaddress}
\author{Moritz Linkmann} 
\email{moritz.linkmann@physik.uni-marburg.de}
\affiliation{\marburgaddress}

\begin{abstract}
The statistical properties of the subgrid energy transfers of  
homogeneous small-scale dynamo are investigated during the kinematic, 
nonlinear and statistically saturated stages.
We carry out an {\em a priori} analysis of data 
obtained from an ensemble of direct numerical simulations on $512^3$ grid points
and at unity magnetic Prandtl number.
In order to provide guidance for subgrid-scale (SGS) modelling of different types
of energy transfer that occur in magnetohydrodynamic dynamos, we consider 
the SGS stress tensors originating from inertial dynamics, Lorentz force and 
the magnetic induction separately. We find that all SGS energy transfers display 
some degree of intermittency as quantified by the scale-dependence of their 
respective probability density functions.  
Concerning the inertial dynamics, a depletion of intermittency occurs in  presence of a saturated dynamo.
\end{abstract}

\maketitle

\section{Introduction}
All turbulent flows are characterized by spatially and temporally chaotic evolutions on a wide range of scales and frequencies \cite{Frisch95}.  As a result,  direct numerical simulations (DNS) approaches are still not practical to study many turbulent flows occurring 
in nature and in engineering applications.  The control parameter is given by the Reynolds number, $Re_L = UL/\nu$ a dimensionless measure of the relative importance of advective and viscous terms in the Navier-Stokes equations (NSE),
where 
$U$ denotes
the rms velocity fluctuations at the energy injection scale, $L$.  It is possible to estimate  that in   homogeneous and isotropic turbulent flows the number of active degrees of freedom grows as $ Re_L^{9/4}$ \ml{\cite{Landau59}}, leading to extremely demanding numerical resources already for moderate turbulent intensities.\\  
To overcome the problem, numerical tools based on a modeling of small-scale turbulent fluctuations are often introduced, and called  large eddy simulations (LES). 
This technique is based on filtering out the small-scale interactions
and replacing them with subgrid-scale (SGS)
models~\cite{Piomelli99,meneveau2000,pope2001}.\\ 
The demand for LES is
increasing for magnetohydrodynamic (MHD) problems, too, as e.g. in  
heliophysical and astrophysical applications~\cite{miesch2015} 
\ml{and in the context of liquid metal MHD. Furthermore, the LES technique 
is a useful alternative to spectral approaches in theoretical analyses of 
interscale energy transfer \cite{Aluie17}, in particular with a view towards
applications in wall-bounded (i.e. liquid metal) flows}.
\ml{In MHD-LES,} the small-scale nonlinear magnetic interactions and the  
velocity/magnetic correlations have to be replaced with SGS models, too.
This introduces additional complexity to the MHD-LES method \cite{Kessar16,Aluie17}, leading to 
different modeling approaches 
\cite{Zhou91,Theobald94,Agullo01,Mueller02a,muller2002,Knaepen04,haugen2006,
Baerenzung08,Balarac10,Chernyshov10}. 
\lb{As in LES of nonconducting fluids, the success of a given  model is usually assessed in terms of reproducing mean profiles of large scale quantities. }
\mb{However, it is more and more clear that SGS velocity fluctuations are characterized by extreme events with magnitudes comparable to that of the large-scale velocity root mean squares. Departure from Gaussian distribution becomes larger and larger by decreasing the scales where velocity and/or magnetic fluctuations are evaluated, a phenomenon known as {\it intermittency} \cite{Frisch95,Biskamp03,Verma04}. As a result, due to their statistical relevance and intensity, extreme events cannot be neglected when modeling SGS dynamics \cite{cerutti1998intermittency,Buzzicotti18a}. Intermittency and anomalous scaling have attracted the attention of several studies on  MHD turbulence \cite{Politano95,Servidio09,Mininni09,Sahoo11,Yoshimatsu11,Imazio13,Meyrand15,Yang17}, with particular interest in high Reynolds number astrophysical applications, e.g. solar wind \cite{Veltri99,Salem09,Wan12,Matthaeus15}. }
The development of SGS models which are sophisticated enough to capture extreme events, 
and therefore provide a more faithful representation of turbulent dynamics, requires
a detailed analysis of  SGS quantities. {\em A-priori} studies of 
DNS data 
provide a first test-bed from where to extract the necessary information. The aim is to  
%
 analyse the SGS correlations of the original fields
and understand what the key features are that must be modeled.  To our knowledge, there are very few {\em a-priori} studies 
for the MHD-LES formulation \cite{Balarac10,Kessar16,Grete16}\ml{\cite{Grete15}}, all of which  concerning  statistically stationary nonlinear dynamos and without any focus on 
intermittency.  
The aim of this paper is to analyse the SGS properties of a MHD turbulent flow
at different temporal instants during the evolution of a small-scale dynamo
such as to be able to assess both regimes, when the magnetic field is passively
advected by, or  actively reacting on, the velocity field. In particular, we
perform a systematic analysis  of the different components of the SGS total
energy transfer. We first split it in two  sub-channels, involving velocity or
magnetic temporal dynamics only and we analyse the mutual scale-by-scale energy
exchanges. Second, we further decompose  
the kinetic SGS energy component into two contributions, one  
coming from the advection and one from the Lorentz force.\\
Furthermore, we also \ml{apply a} 
formulation of the filtered fields,
based on an exact projection on a finite number of Fourier modes 
(P-LES) \cite{Buzzicotti18a} that disentangles the signal due to the  
coupling between resolved and unresolved scales from that 
due to interactions between resolved fields only. \\
     
The main results of this study are: \\
\ml{ (i) The SGS energy transfer shows some degree of intermittency in all
evolutionary stages of the dynamo. Its component coming from the Lorentz force
becomes successively more intermittent while that originating from
hydrodynamics shows decreased intermittency. \\ (ii) In terms of guidance for
LES modelling, we find that dissipative models should be well suited for the
SGS stresses connected with the Lorentz force, while not being suitable for
those coming from purely inertial dynamics.  }

This paper is organized as follows: We begin in section \ref{sec:dataset} 
with a description of
the DNS dataset.  In section \ref{sec:theory}, we introduce the P-LES formulation for MHD.
Section \ref{sec:apriori} presents the results from the {\em a-priori} analysis of the
statistical properties of the SGS energy transfers. We summarize and discuss
our results in section \ref{sec:conclusions}.

\section{Description of the dataset}
\label{sec:dataset}

\begin{table}[t]
  \centering
  \begin{tabular}{ccccccccc}
    \ml{$M$} & $\Re_L$ & $\Rl$	& $\varepsilon$ & $U$ & $L$ & $\nu$ & $T$  & $k_{max}\eta_u$	\\
    \hline
    512 & 889 & 164	& 0.14	& 0.61	& 1.0	& 0.0007 & 1.7 & 1.3	\\
    \hline
  \end{tabular}
  \label{tab:dataset}
  \caption{Description of the statistically stationary hydrodynamic simulation used as
          an initial condition for the velocity fields in the dynamo runs.
   \ml{$M$} denotes the number of grid points in each Cartesian coordinate,
   \ml{$\Re_L$ the integral-scale Reynolds number, $\Rl$ the Reynolds number with respect to the Taylor microscale}, 
   $\varepsilon$ the total
   dissipation rate, $U$ the rms velocity, $L$ the integral length scale of the
   turbulence, $\nu$ the kinematic viscosity, 
   $T=L/U$ the large-eddy turnover time, \ml{$k_{\rm max}$ the \ml{largest} 
    resolved wave number and $\eta_u$ the Kolmogorov microscale.} All observables are time averaged. 
}
\end{table}

\noindent
The data for the \textit{a priori} study is generated through DNSs of 
the three-dimensional incompressible MHD equations
\begin{align}
  \p{t}\bu + \left(\bu \cdot \nabla\right)\bu &= -\nabla p + \left(\nabla \times \bb\right) \times \bb + \nu \Delta \bu + \vec{f}, \label{eq:mom}\\
  \p{t}\bb &= \nabla \times \left(\bu \times \bb\right) + \eta \Delta \bb, \label{eq:ind}\\
  \nabla \cdot \bu & = 0, \ \ 
  \nabla \cdot \bb  = 0, \label{eq:sol}
\end{align}
where $\vec{u}$ is the velocity field, $\vec{b}$ the magnetic field in Alfv\'en units, 
$p$ the pressure {divided by the density}, $\nu$ the kinematic viscosity, 
$\eta$ the magnetic resistivity, and 
$\vec{f}$ an external mechanical force which is solenoidal at all times. 
The density has been set to unity for convenience. 

Equations \eqref{eq:mom}-\eqref{eq:sol} are solved numerically on the periodic domain 
$V = [0,2\pi]^3$ using the 
pseudospectral method \cite{Orszag69} with full dealiasing by the $2/3$rds rule \cite{Orszag71}.
An ensemble of 10 runs is generated, where the initial velocity field configurations 
are obtained from a statistically stationary hydrodynamic 
DNS on $512^3$ grid points by sampling in intervals of one large-eddy turnover time
\ml{$T=L/U$, where $U$ is the rms velocity and $L$ the integral scale of the 
turbulence.
The mechanical force $\vec{f}$ is a Gaussian-distributed and delta-in-time correlated random process acting at wavenumbers $1 \leq k \leq 2.5$ with a flat spectrum and without injection of kinetic helicity.  
The magnetic seed fields are randomly generated with a Gaussian distribution and concentrated at wavenumber $k_{\rm s}=40$.}
Details of the stationary hydrodynamic simulation are summarized in table I.

\begin{table*}[t]
	\begin{tabular}{cccccccccccccc}
		& $\Re_L$ & $\Rl$ & \ml{$\Pm$} & $\varepsilon_u$ & $\varepsilon_b$ & $U$ & $L$ & $B$ & $L_b$ & \ml{$\eta$} & $k_{\rm max} \eta_u$ & $k_{\rm max} \eta_b$ & $t_S/T$  \\
		\hline
		(I) & 811 & 161 & 1 & 0.099 & $2.6\cdot 10^{-3}$ & 0.59 & 0.97 & 0.020 & 0.092 & 0.0007 & 1.3  & 3.2  & 8.8 \\
		(II) & 851 & 208 & 1 & 0.057 & 0.056 & 0.58 & 1.0 & 0.13 & 0.15 & 0.0007 & 1.5  & 1.5  & 17.6 \\
		(III) & 870 & 211 & 1 & 0.032 & 0.076 & 0.51 & 1.2 & 0.25 & 0.29 & 0.0007 & 1.7  & 1.4 & 32.3 \\
		\hline
	\end{tabular}
	\label{tab:stage_properties}
	\caption{
         Summary of the dynamo simulations during kinematic (I), nonlinear (II) and saturated stages (III).
         $\Re_L$ denotes the integral-scale Reynolds number, $\Rl$ the Reynolds number with respect to the 
         Taylor microscale, \ml{$\Pm$ the magnetic Prandtl number}, $\varepsilon_u$
         the kinetic dissipation rate, $\varepsilon_b$ the magnetic dissipation rate, $U$
         the rms velocity, $L$ the integral length scale of the turbulence, $B$ the rms of the magnetic
         field, $L_b$ the magnetic integral length scale, \ml{$\eta$ the resistivity}, 
         $k_{\rm max}$ the \ml{largest} 
         resolved wavenumber, $\eta_u$ and $\eta_b$ are the kinetic and magnetic Kolmogorov microscales,
         respectively, and $t_S$ is the sampling time of each evolutionary stage of the dynamo.
         All observables are ensemble-averaged over an ensemble of 10 simulations.
         }
\end{table*}

\begin{figure}[t]
  \centering
  \includegraphics[width=0.5\textwidth]{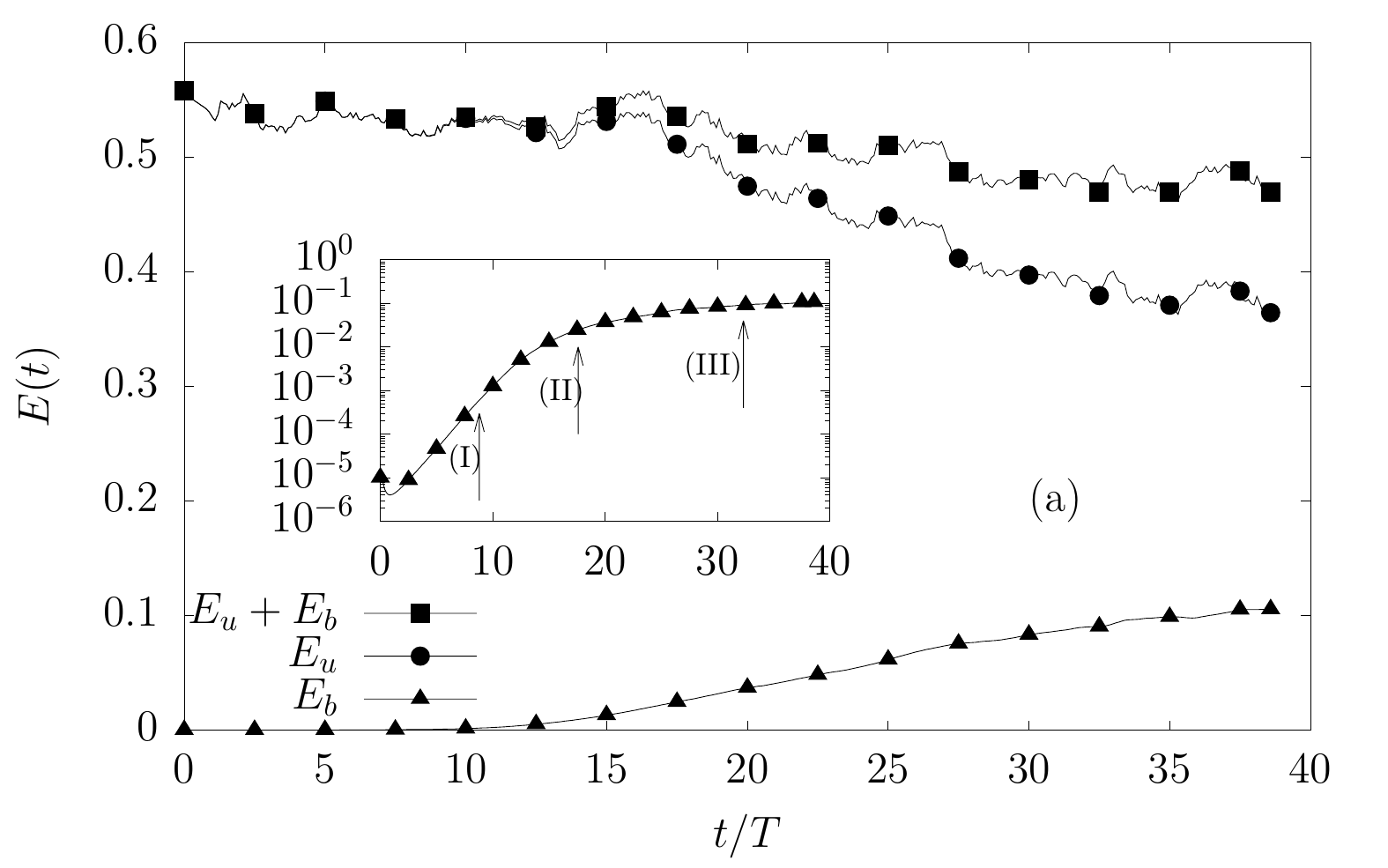}
  \includegraphics[width=0.5\textwidth]{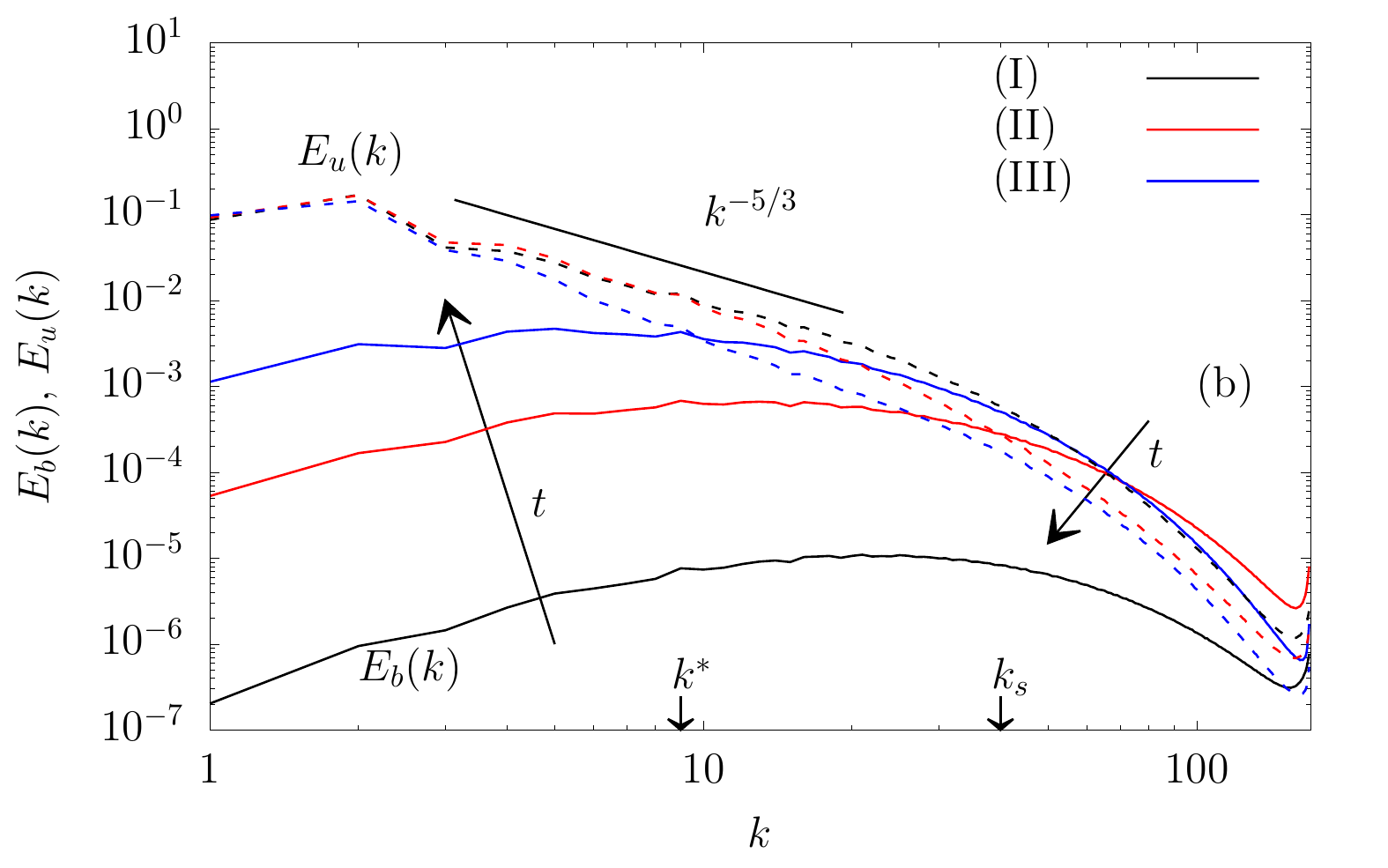}
  \caption{
   Panel (a): 
   Time evolution of the kinetic energy $E_u$, the magnetic energy 
   $E_b$ and the total energy $E_u+E_b$ with time measured in units of 
   large-eddy turnover time $T$ (see table 1). 
   The inset shows the evolution $E_b$ on a
   linear-logarithmic scale to highlight its initial exponential growth 
   phase. The different stages of dynamo evolution are indicated
   by arrows: (I): kinematic stage, (II) nonlinear stage, (III) saturated stage.
   Panel (b): Kinetic energy spectra $E_u(k)$ (dashed) and magnetic energy spectra 
   $E_b(k)$ (solid) measured at $t/T =8.8$ in stage (I), at $t/T =17.6$ 
   in stage (II) and at $t/T = 32.3$ in stage (III).}
  \label{fig:En_evo}
\end{figure}

The time evolution of the kinetic and magnetic energies 
per unit volume
\begin{align}
E_u(t) &= \frac{1}{2} \left \langle |\vec{u}(\vec{x},t)|^2 \right \rangle_{{V,N}} 
        \equiv \left \langle \frac{1}{2|V|}\int d\vec{x} \ |\vec{u}(\vec{x},t)|^2 \right \rangle_{{N}} \ , \\
E_b(t) &= \frac{1}{2} \left \langle |\vec{b}(\vec{x},t)|^2 \right \rangle_{{V,N}} 
        \equiv \left \langle \frac{1}{2|V|}\int d\vec{x} \ |\vec{b}(\vec{x},t)|^2 \right \rangle_{{N}} \ , 
\end{align}
{where the subscript $N$ denotes an ensemble average over $N=10$ realisations,}
and the total energy for the ensemble are shown in Fig.~\ref{fig:En_evo}(a).
From the time evolution of $E_b$, which is also shown on a 
linear-logarithmic scale in the inset, it can be seen that
the simulation can be divided in three stages. 
First, during the kinematic stage (I), the magnetic field 
grows exponentially. During that stage the Lorentz force {in Eq.~\eqref{eq:mom}} is negligible and 
the evolution equations are linear in the magnetic field. The exponential growth phase 
ends once the Lorentz force 
is large enough such that the back-reaction of the magnetic field on the flow needs to 
be taken into account. 
This is the nonlinear, unsteady, stage (II) of the evolution, during which 
$E_b$ continues to increase sub-exponentially \ml{\cite{Haugen04,Schekochihin04}}.  
Finally, $E_b$ is approaching a statistically stationary
state. \ml{That is,} it enters the saturated stage (III)
\ml{which at unity magnetic Prandtl number and sufficiently large $\Re$ 
is characterized by the ratios of the dissipation rates
$\varepsilon_b/(\varepsilon_u + \varepsilon_b) \simeq 0.7$ and energies $E_b/(E_u+E_b) \simeq 0.25$ \cite{Haugen03,Brandenburg14,Linkmann17,McKay17}. Our data in stage (III)
is consistent with these ratios, as can be seen from the values listed in 
table \ref{tab:stage_properties}, where a summary of the dynamo runs in the kinematic (I), nonlinear (II) and saturated (III) stages is provided.} 
The SGS energy transfers will be studied during stages (I)-(III), with each stage
analysed separately.

The kinetic and the magnetic energy spectra 
\begin{align}
E_u(k,t) & = \frac{1}{2} \left \langle \int_{|\vec{k}|=k} |\fvec{u}_{\vec{k}}(t)|^2  \ d\vec{k} \right \rangle_N \ , \\ 
E_b(k,t) & = \frac{1}{2} \left \langle \int_{|\vec{k}|=k} |\fvec{b}_{\vec{k}}(t)|^2 \ d\vec{k} \right \rangle_N \ ,  
\end{align}
are shown in Fig.~\ref{fig:En_evo}(b) for different instances in time corresponding to 
stages (I)-(III) as specified in table \ref{tab:stage_properties}. 
The kinetic energy spectrum is dominated by the forcing in the interval $1\leq k\leq2.5$. 
During the kinematic stage, an inertial subrange with Kolmogorov scaling can be identified, 
as indicated in the figure by the straight solid line. During stages (II) and (III) we observe a 
steepening of $E_u(k)$ at successively smaller wavenumbers.   
The magnetic energy spectrum grows self-similarly during {stage (I)} 
which is typical for a small-scale dynamo~\cite{Schekochihin04, Mininni05a, Brandenburg05}.
In the saturated stage (III), the magnetic energy exceeds the kinetic energy
at the small scales while the large scales remain essentially hydrodynamic and 
forcing-dominated. A crossover-wavenumber $k^*$ can be identified where $E_u(k^*) = E_b(k^*)$, 
in the present dataset $k^* = 9$.
\ml{Since the peak of the saturated magnetic energy spectrum depends on the forcing scale \cite{Brandenburg05}, the equipartition scale we measure will also not be universal, that is, it should depend on the forcing.  
}

\section{P-LES formulation for MHD}
\label{sec:theory}
The governing equations 
are derived by applying a filtering operation to the MHD equations \cite{Zhou91,Kessar16,Yang16a,Aluie17}, {with}
the filtered {component} 
\ml{$\overline h$ of a function $h$} 
defined as
\begin{align}
  \overline{h}(\vec{x},t) \equiv 
	\int_{V}d\vec{y}\, G\left(\vec{x} - \vec{y}\right) h(\vec{y},t) = 
  \sum\limits_{\vec{k} \in \mathbb{Z}^3} \hat{G}(\vec{k}) \hat{h} (\vec{k},t) e^{i\vec{k}\vec{x}},
  \label{eq:filter}
\end{align}
where $G$ is the filter function and $\hat{\cdot }$ denotes the Fourier transform.  Applying
this filtering operation to Eqs.~\eqref{eq:mom} and \eqref{eq:ind}, we obtain 
the filtered momentum and induction equations 
given here in tensor notation
\begin{align}
  \p{t}\ou_i &= - \p{j}\left(\overline{\ou_i\ou_j} - \overline{\ob_i\ob_j} + \tij{I} - \tij{M} + \overline{p}\delta_{ij}\right) + \nu \p{jj} \ou_i + {\overline{f}_i}, 
  \label{eq:momentum_PLES}\\
	\p{t}\ob_i &= - \p{j}\left(\overline{\ob_i\ou_j} - \overline{\ou_i\ob_j} + \ml{\tij{b}} \right) + \eta \p{jj} \ob_i,
  \label{eq:induction_PLES}
\end{align}
{where a summation over repeated indices is implied, and}
with 
\begin{align}
  \tij{I} =& \ \overline{u_i u_j} - \overline{\ou_i \ou_j}, \label{eq:SGS_tensor_V} \\
  \tij{M} =& \ \overline{b_i b_j} - \overline{\ob_i \ob_j}, \label{eq:SGS_tensor_M} \\
  \tij{b} =& \ \overline{b_i u_j} - \overline{\ob_i \ou_j} 
	           - (\overline{u_i b_j} - \overline{\ou_i \ob_j}), \label{eq:SGS_tensor_b} 
\end{align}
where $\tij{I}$ is the inertial SGS tensor, $\tij{M}$ the Maxwell SGS tensor,
\ml{and $\tij{b}$ the SGS tensor originating from the electromotive force 
in Eq.~\eqref{eq:ind}.}
\ml{
It consists of two SGS-stresses which} 
are related to each other by transposition. 
They are associated with different dynamics, 
that is with advection, $(\bu\cdot\nabla)\bb$, \ml{in case of $\overline{b_i u_j} - \overline{\ob_i \ou_j}$} 
or dynamo action through magnetic field-line stretching, $(\bb\cdot\nabla)\bu$, 
\ml{in case of $\overline{u_i b_j} - \overline{\ou_i \ob_j}$}. 
\ml{However, as} they have a common physical origin, 
the electric field $\vec{E} = \bu \times \bb$, 
\ml{we do not consider them separately}. 

{Equations \eqref{eq:momentum_PLES}-\eqref{eq:induction_PLES}} are obtained by
using \ml{solenoidality} 
of both fields, the linearity of the filtering operator
and including the terms which can be written as a gradient into the pressure gradient.
\ml{As usual, the
equations are not closed in terms of the resolved fields only, due to the fact
that the SGS stress tensors depend on the product of two unresolved fields.}
\ml{Equations ~\eqref{eq:momentum_PLES}-\eqref{eq:induction_PLES}
differ from those usually given in the MHD literature on LES \cite{Zhou91,Kessar16,Aluie17}
through the additional filtering of products of two resolved fields. 
In conjunction with a projector filter,
the latter ensures that after introducing SGS models, 
Eqs.~\eqref{eq:momentum_PLES} and \eqref{eq:induction_PLES} 
can be evolved on a finite computational grid \cite{Sagaut06, Buzzicotti18a}, 
which can seen by supposing that $G$ in Eq.~\eqref{eq:filter}
is a Galerkin projector on a finite number of 
Fourier modes \cite{Buzzicotti18a}.  
In what follows, we consider $G$ to be a projector and 
Eqs.~\eqref{eq:momentum_PLES}-\eqref{eq:SGS_tensor_b}
are referred to as the P-LES formulation.
}
%

%
%

The P-LES formulation has the further advantage that, unlike in the \ml{usual} 
LES formulation, the SGS energy transfers based on the P-SGS tensors 
\ml{defined in Eqs.~\eqref{eq:SGS_tensor_V}-\eqref{eq:SGS_tensor_b}} 
do not contain couplings between the resolved fields \cite{Buzzicotti18a}. 
The latter is very important for the evaluation of backscatter in {\em a priori} analyses of SGS energy
transfers, since residual couplings between the resolved fields can be wrongly interpreted 
as backscatter events. We will come back to this point in Secs.~\ref{sec:energy_transfers} 
and \ref{sec:energy_transfers-tot}.

\blue{
Finally, we point out that care must be taken in {\em a-posteriori} 
studies of MHD LES concerning $\overline{p}$ since it contains the magnetic 
SGS pressure term, which is not closed in terms of the resolved magnetic field. 
As such, a closure of Eq.~\eqref{eq:SGS_tensor_M} would lead to two models 
for the magnetic pressure term: an explicit one coming from the choice of model 
and an implicit one from the solution of the Poisson equation. 
However, the magnetic pressure term does not affect the global energy transfers
and is thus not of direct relevance to the present {\em a-priori} study.
}

\subsection{The resolved-scale energy transfer}
\label{sec:energy_transfers}
Neglecting viscous, Joule dissipation {and  forcing} terms, 
the {P-LES} kinetic and magnetic energy evolution equations read
\begin{align}
  \p{t}\frac{1}{2}\ou_i\ou_i + \p{j}A^{u}_j &= -\Pi^{u} + (\p{j}\ou_i) (\overline{\ou_i \ou_j}) - (\p{j}\ou_i) (\overline{\ob_i \ob_j}), \label{eq:kin_en}\\
  \p{t}\frac{1}{2}\ob_i\ob_i + \p{j}A^{b}_j &= -\Pi^{b} + (\p{j}\ob_i) (\overline{\ob_i \ou_j}) - (\p{j}\ob_i) (\overline{\ou_i \ob_j}), \label{eq:mag_en}
\end{align}
where $A_j^{u} = \ou_i(\overline{\ou_i \ou_j} - \overline{\ob_i \ob_j} +
\overline{p} \delta_{ij} + \tij{I} - \tij{M})$ and  $A_j^{b} =
\ob_i(\overline{\ob_i \ou_j} - \overline{\ou_i \ob_j} +\ml{\tij{b}})$ 
{result in flux terms}
that redistribute the energies in space and vanish under
spatial averaging: 
{$\langle \p{j}A_j^{u}\rangle_V = \langle \p{j}A_j^{b}\rangle_V = 0$ }.  
The {P-SGS} 
energy transfers $\Pi^{u}$ and $\Pi^{b}$ are defined as
\begin{align}
  \Pi^{u} &= \Pi^{I} - \Pi^{M} = (\p{j} \ou_i) \tij{I} -  (\p{j} \ou_i) \tij{M},\\
  \Pi^{b} &= \ml{(\p{j} \ob_i) \tij{b}},
\end{align}
where $\Pi^{I} = -(\p{j}\ou_i)\tij{I}$ is the inertial SGS energy transfer,
$\Pi^{M} = -(\p{j}\ou_i)\tij{M}$ the Maxwell SGS energy transfer
\ml{and $\Pi^{b}$ the SGS energy transfer associated with 
the electromotive force.}
Equations~\eqref{eq:kin_en} and \eqref{eq:mag_en} contain four extra terms:
$(\p{j}\ou_i) (\overline{\ou_i \ou_j})$, $(\p{j}\ou_i) (\overline{\ob_i
\ob_j})$, $ (\p{j}\ob_i) (\overline{\ob_i \ou_j})$ and $(\p{j}\ob_i)
(\overline{\ou_i \ob_j})$. 
\ml{Using $\nabla \cdot \overline{\vec{u}} =0$,
$\nabla \cdot \overline{\vec{b}} =0$ 
and the projector property $\hat{G}^2 = \hat{G}$}, it can be shown that 
\begin{align}
& \avg{(\p{j}\ou_i)(\overline{\ou_i \ou_j})}_V = 0 \ , \\  
&\avg{(\p{j}\ob_i) (\overline{\ob_i \ou_j})}_V = 0 \ ,
\end{align}
and 
\begin{equation}
\avg{(\p{j}\ou_i) (\overline{\ob_i \ob_j})}_V = - \avg{(\p{j}\ob_i) (\overline{\ou_i \ob_j})}_V \ , 
\end{equation}
{hence} they do not contribute to the global {total energy} balance.
Furthermore, it is easy to verify that out of the four terms 
only $(\p{j}\ou_i) (\overline{\ob_i\ob_j})$ is Galilean invariant. 
Galilean invariance is important to prevent the occurrence 
of unphysical fluctuations in the measured SGS energy transfer 
\cite{aluie2009I,aluie2009II,Buzzicotti18a}.  
This problem can be solved by adding and subtracting energy transfers 
originating from the Leonard stress components for each SGS 
tensor~\cite{leonard1975,Zhou91} in Eqs.~\eqref{eq:kin_en} and 
\eqref{eq:mag_en}. 
{The Leonard stresses are defined as}
\begin{align}
  \tij{I,L} & = \overline{\ou_i \ou_j} - \ou_i \ou_j,\\
  \tij{M,L} & = \overline{\ob_i \ob_j} - \ob_i \ob_j,\\
  \tij{b,L} & = \ml{\overline{\ob_i \ou_j} - \ob_i \ou_j -(\overline{\ou_i \ob_j} - \ou_i \ob_j)}, 
\end{align}
{which give rise to the following energy transfer terms}
\begin{align}
  \label{eq:Leo_u}
	\Pi^{u,L} & = \Pi^{I,L} - \Pi^{M,L} = (\p{j} \ou_i) \tij{I,L} -  (\p{j} \ou_i) \tij{M,L},\\
  \label{eq:Leo_b}
	\Pi^{b,L} & = \ml{(\p{j} \ob_i) \tij{b,L}} .
\end{align}
{Including the Leonard terms in Eqs.~\eqref{eq:kin_en} and \eqref{eq:mag_en}} results in 
\begin{align}
  \label{eq:evol_Eu}
  \p{t}\frac{1}{2}\left(\ou_j \ou_j\right) & + \p{j}\left(A^{u}_j + \ou_i\tij{u,L}\right) \nonumber \\
                                           &= - \Pi^{u} - \Pi^{u,L} - (\p{j}\ou_i) (\ob_i \ob_j),\\
  \label{eq:evol_Eb}
  \p{t}\frac{1}{2}\left(\ob_j \ob_j\right) &+ \p{j}\left(A^{b}_j + \ou_i\tij{b,L}\right) \nonumber \\
                                           &= - \Pi^{b} - \Pi^{b,L} + (\p{j}\ou_i) (\ob_i \ob_j).
\end{align}
Now all terms in the resolved energy evolution equations are Galilean invariant.\\

\noindent
It is important to remark that the Leonard SGS transfers vanish under spatial averaging,
i.e. they do not alter the global balances. {Furthermore, they couple only the 
resolved fields, hence they cannot be associated with transfers between resolved and 
SGS quantities. Therefore the LES formulation differs from the P-LES formulation
in a fundamental way: All SGS-tensors in the LES formulation are the sum of the 
respective P-SGS and Leonard tensors, e.~g. $\tij{I,\rm LES} = \tij{I} + \tij{I,L}$, 
and the corresponding SGS energy transfers of the LES formulation contain 
the contribution from the Leonard stresses. That is, the SGS energy transfers
in the LES formulation have contributions from interactions
between the resolved fields \cite{Buzzicotti18a}. We will come back to this
point in the context of backscatter in Sec.~\ref{sec:energy_transfers-tot} and in the Appendix.} 

\begin{figure}[htp]
	\centering
	\includegraphics[width=0.45\textwidth]{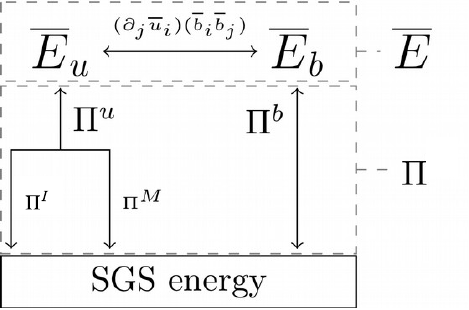}
	\caption{
         A schematic representation of the energy transfer between the 
         resolved-scale energies and the SGS energy. The exchange between
         magnetic and kinetic energies at the resolved scales is carried
         by the resolved-scale conversion term 
         $(\p{j}\ou_i)\ob_i\ob_j$. {According to Eqs.~\eqref{eq:evol_Eu} and \eqref{eq:evol_Eb},} the exchange 
         of energy between resolved scales and SGS follows different channels, 
         $\Pi^{b}$ couples the resolved-scale magnetic energy to the SGS and combines 
         the physical processes of advection of magnetic energy and magnetic field 
         line stretching, while $\Pi^{u}$ transfers kinetic energy between resolved scales and SGS. 
         The latter itself has two components, an inertial channel $\Pi^{I}$ which is due to 
         vortex-stretching and advection and a magnetic channel $\Pi^{M}$ originating from 
         the Lorentz force. During the kinematic stage of the dynamo, $\Pi^{M}$ is negligible
         compared to $\Pi^{I}$. 
        }
	\label{fig:diagram}
\end{figure}

\noindent
{Finally, the term $(\p{j}\ou_i) (\ob_i \ob_j)$ occurs in Eqs.~\eqref{eq:evol_Eu} and \eqref{eq:evol_Eb}} with opposite sign. 
Since it 
is closed in terms of the resolved fields and exchanges kinetic and magnetic energy, 
{$(\p{j}\ou_i) (\ob_i \ob_j)$} has been named 
{\em resolved-scale conversion term}~\cite{Aluie17}. 
\ml{It is positive if kinetic energy is converted to magnetic energy and 
negative vice versa.}
%
With $\Pi^{u} = \Pi^{I} - \Pi^{M}$ and $\Pi^{b}$ 
we now have key benchmark quantities to study the properties of the
different SGS energy transfers. 
Furthermore, as the total energy is conserved in the absence of forcing and dissipation, 
the total SGS energy transfer is also a quantity of interest.
We define the resolved total energy transfer \ml{$\Pi$  through the resolved-scale total energy balance} 
\begin{align}
  \p{t}\frac{1}{2}\left(\ou_i \ou_i\right) + \p{t}\frac{1}{2}\left(\ob_i \ob_i\right) &+  \p{j}\left(A_j + \ou_i\tij{L}\right) \nonumber \\ 
     &= - \Pi - \Pi^L
  \label{eq:total_ples}
\end{align}
where $A_j = A^u_j + A^b_j$, $\tij{L} = \tij{u,L} + \tij{b,L}$, $\Pi = \Pi^{u} + \Pi^{b}$ and $\Pi^L = \Pi^{u,L} + \Pi^{b,L}$.
Figure~\ref{fig:diagram} gives a schematic overview of the different 
SGS energy transfers.

\section{\textit{A priori} analysis of the SGS energy transfers}
\label{sec:apriori}

The \textit{a priori} analysis of the statistical properties of the SGS energy
transfers is carried out using a sharp spectral cut-off filter, 
\ml{which is 
defined through its action on a generic \ml{function $h$} 
\begin{align}
  \overline{h}(\vec{x},t) \equiv 
  \sum\limits_{|\vec{k}| < k_c} \hat{h} (\vec{k},t) e^{i\vec{k}\vec{x}},
  \label{eq:galerkin_filter}
\end{align}
where $k_c$ is the cut-off wavenumber, which corresponds to 
the configuration-space filter width $\Delta = \pi/k_c$.}
\mb{Although sharp projectors produce
Gibbs oscillations in \ml{configuration} 
space \cite{ray2011resonance} resulting in 
SGS stress tensors \cite{vreman1994realizability}
\ml{that are not positive-definite}, they have
the advantage to create a clear distinction between resolved and unresolved
scales and to allow all terms in the equations
\eqref{eq:momentum_PLES}-\eqref{eq:induction_PLES} to evolve on the same
Fourier subspace for all times.  Moreover, for nonconducting flows, a good
agreement between the statistics of the SGS energy transfer obtained from a
sharp cutoff and Gaussian filter was found \cite{Buzzicotti18a}, suggesting
that 
\ml{effects specific to Galerkin projection have}
only a subleading effect at the level of the energy evolution equations. 
} \\

In what follows, we study the mean ({here, mean refers to the combined spatial and ensemble average})  
{P-}SGS energy transfers 
and their spatial fluctuations for different $k_c$.  
\mb{The fluctuations are investigated through the probability density functions (pdfs)
of the respective P-SGS energy transfers. 
\ml{I}n order to quantify the departure from Gaussianity at 
different scales, it is customary to evaluate the flatness $F_x$ of the standardized pdfs, 
$$
F_x(k_c) = \langle x^4 \rangle/ \langle x^2 \rangle^2 \sim k_c^\zeta, 
$$
as a function of the cutoff wavenumber, where $x$ represents the different contributions of the P-SGS energy transfer.}
Since the Leonard stresses do not provide information relevant to modelling, 
we summarize results specific to the Leonard stresses in the Appendix, 
which is referenced in the text where necessary.

\noindent
We begin with 
$\Pi$,
and subsequently increase the level of detail by first splitting 
$\Pi$ into 
$\Pi^{u}$ and $\Pi^{b}$, 
{followed by the decomposition of} 
$\Pi^{u}$ into 
$\Pi^{I}$ and $\Pi^{M}$.  
Note that $\Pi^{b}$ is not decomposed any further, because
the stress tensors \ml{associated with the advection and field-line stretching terms 
in the induction equation} 
originate both from the electric field and are related to each other 
by transposition, as discussed in Sec.~\ref{sec:theory}. As such, a single LES
model term should be used in the induction equation.

\subsection{The total SGS energy transfer} 
\label{sec:energy_transfers-tot}

\begin{figure}[t]
  \centering
  \includegraphics[width=0.5\textwidth]{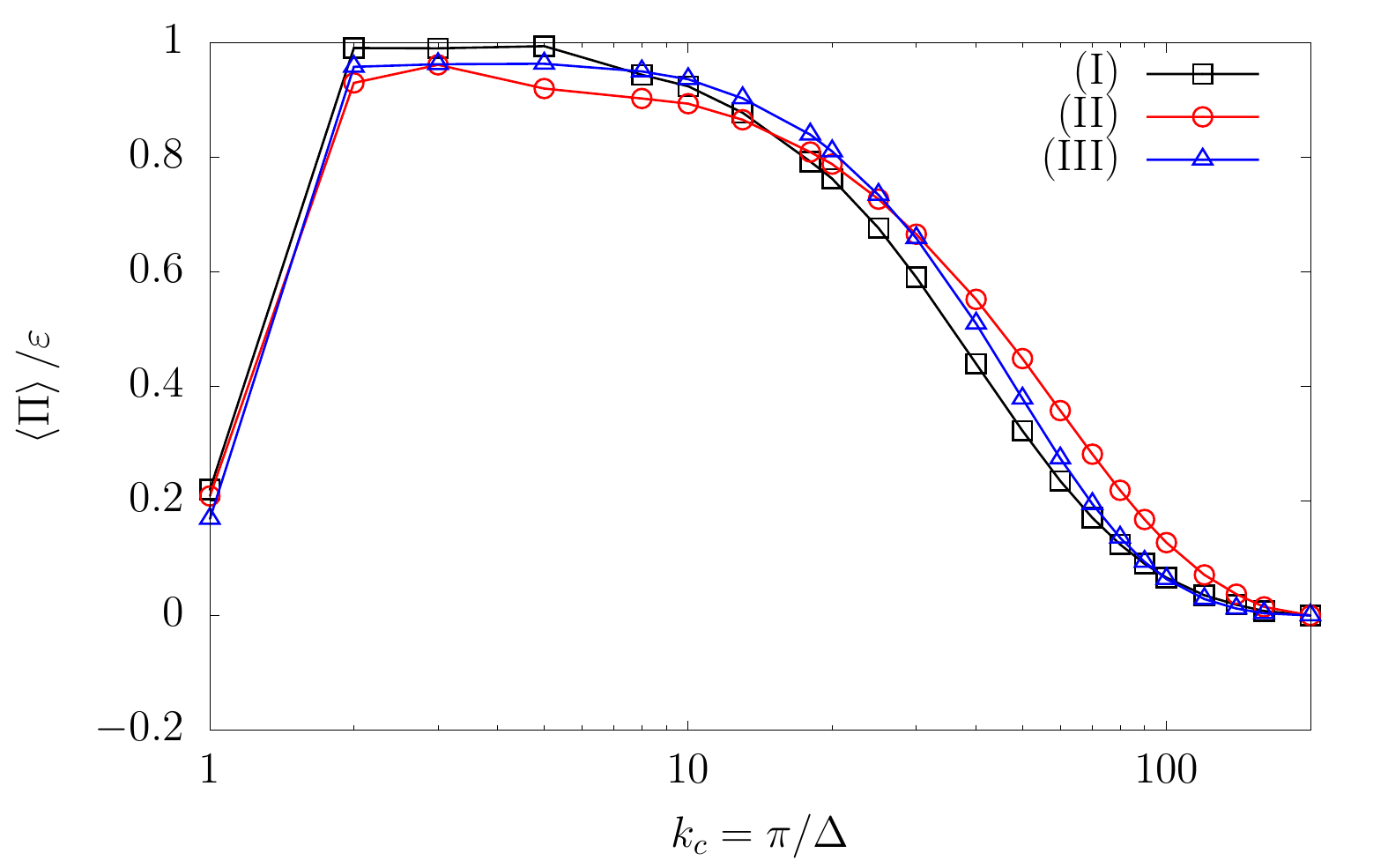}
  \caption{
	The mean 
    total SGS energy transfer $\langle \Pi\rangle_{V,N} $ normalized
	with the total dissipation $\varepsilon$ versus the cutoff wavenumber $k_c$ at the
	kinematic (I), nonlinear (II) and stationary (III) stages.}
  \label{fig:TOT_mean}
\end{figure}

Figure \ref{fig:TOT_mean} presents $\avg{\Pi}_{V,N}$ as function of $k_c$ 
at three different instants during the time evolution which are representative
of the three stages (I)-(III).
\mb{Since $\Pi$ is obtained using a spectral cut-off projector, its mean value equals the total energy flux in Fourier space across the cut-off wavenumber $k_c = \pi/\Delta$, see  \cite{eyink2005locality,Buzzicotti18a}.}
As can be seen from Fig.~\ref{fig:TOT_mean}, 
$\avg{\Pi}_{V,N} \geqslant 0$, which is representative of a mean total energy transfer 
from large scales to small scales. Furthermore, we find that $\avg{\Pi}_{V,N}$
does not change significantly during the different evolutionary stages of the 
dynamo, which implies that the exchange of kinetic and magnetic energy proceeds
in a way that leaves the total scale-by-scale transfer unaffected. We will 
come back to this point in further detail when assessing the decomposed
SGS energy transfers.  

\begin{figure*}[t]
  \centering
  \includegraphics[width=0.45\textwidth]{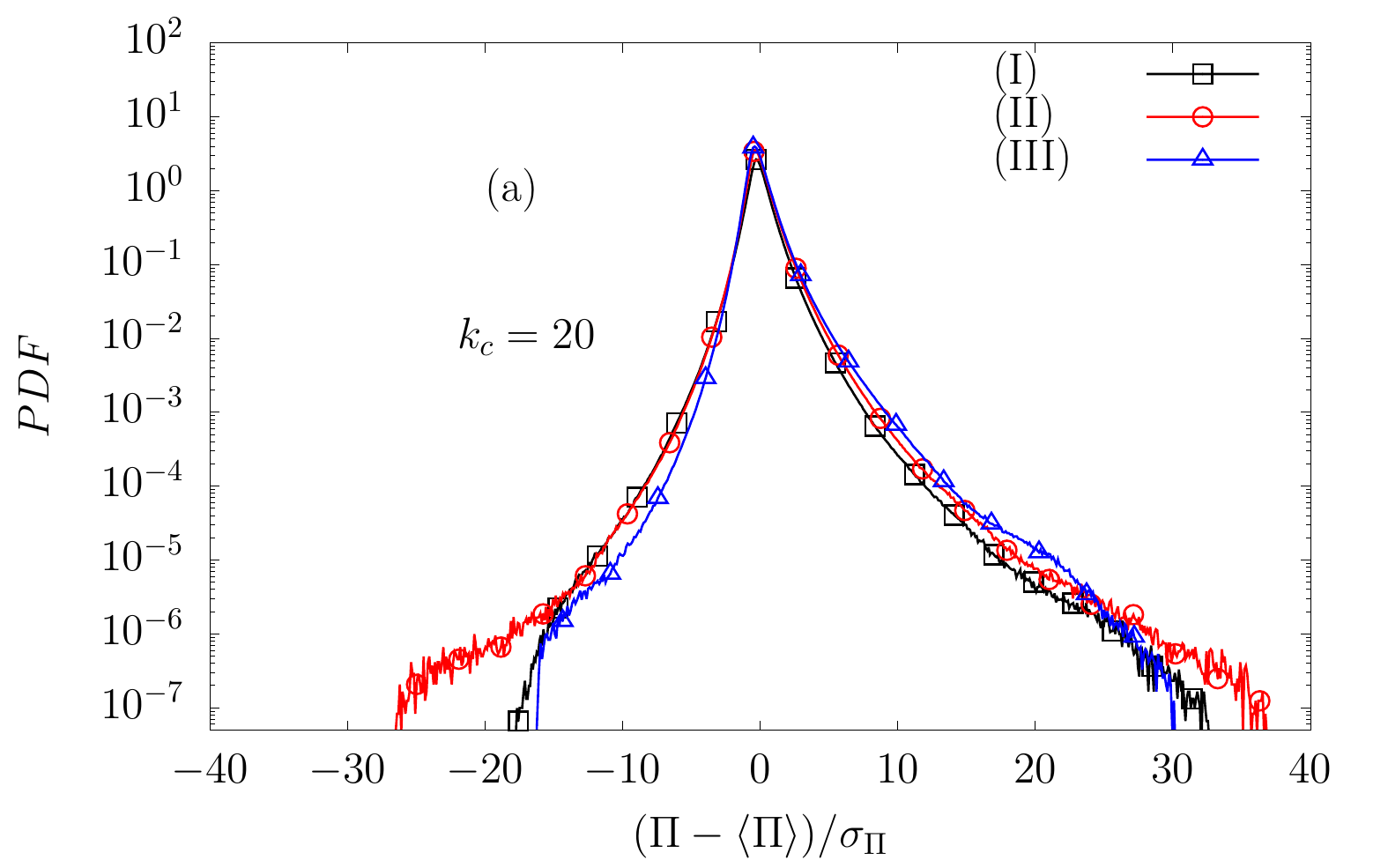}
  \includegraphics[width=0.45\textwidth]{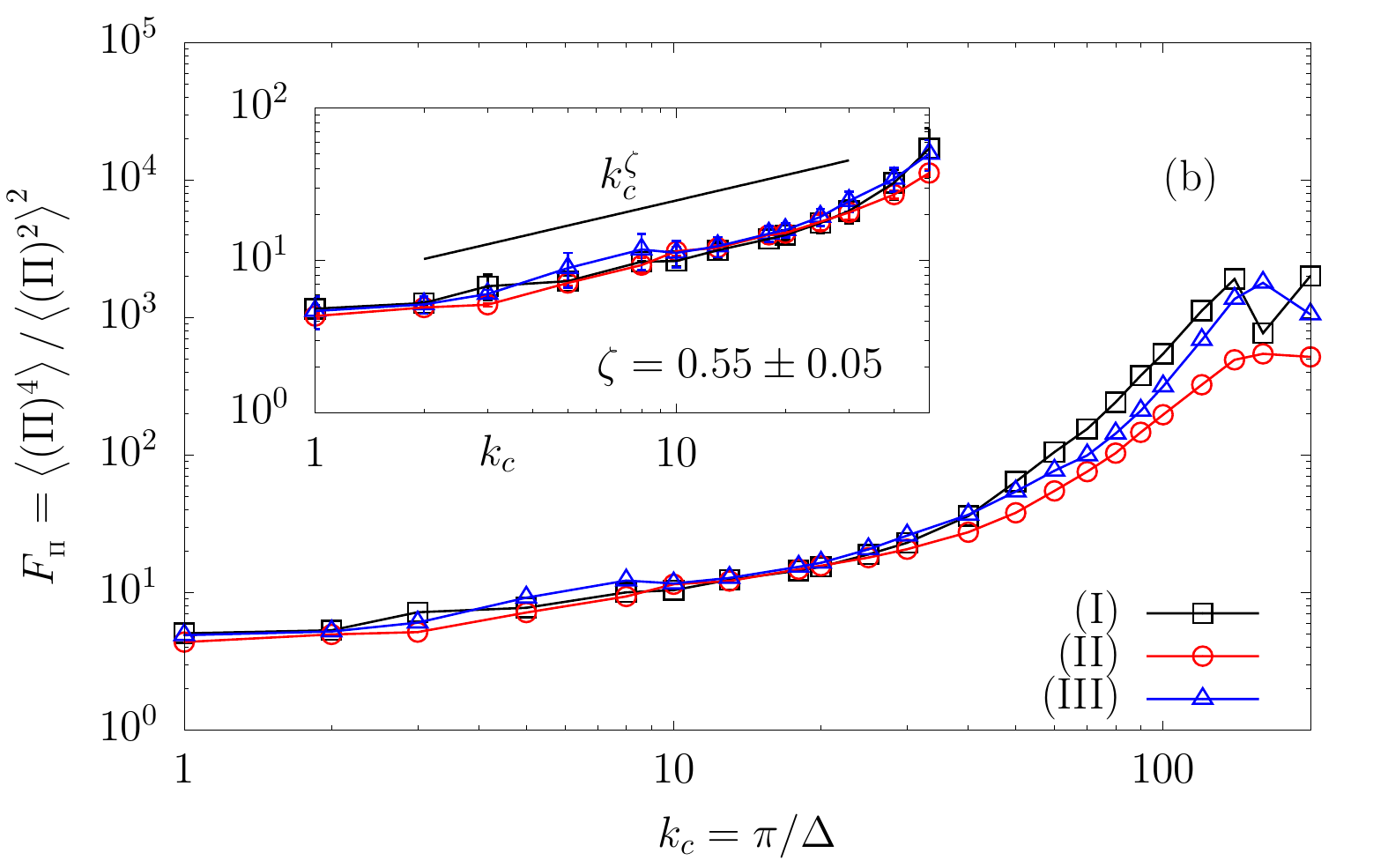}
  \includegraphics[width=0.45\textwidth]{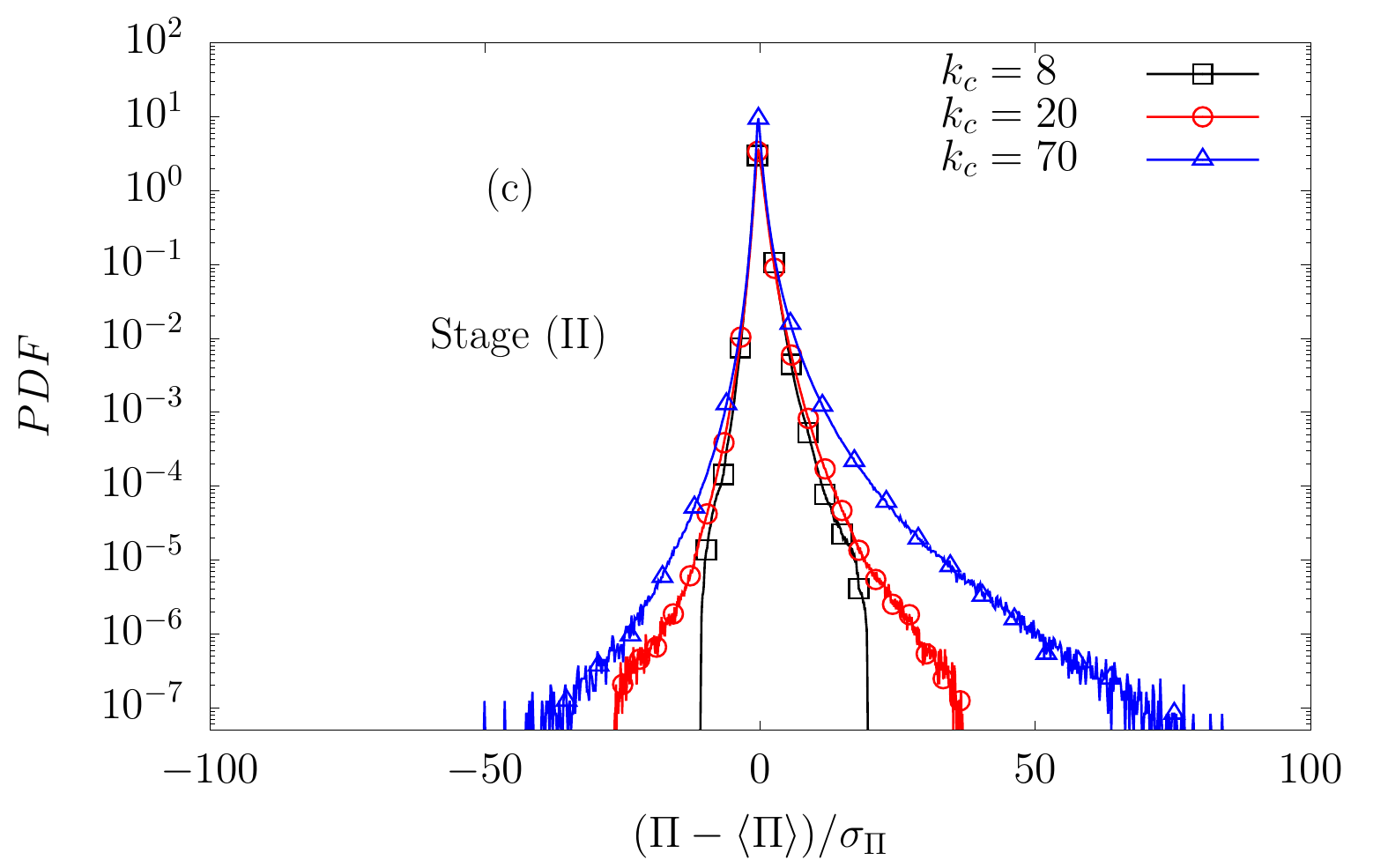}
  \caption{Total SGS energy transfer $\Pi$ during the kinematic stage (I), non-linear stage (II) and the stationary stage (III).
	Panel (a): pdfs of $\Pi$ at $k_c=20$.
	Panel (b): The flatness of $\Pi$ versus the
	cutoff wavenumber $k_c$. \mb{The inset shows a zoom of the flatness in the inertial range of scales, with error bars estimated from the different configurations. The black line represents the fit of the flatness scaling exponent in the range $2 \leqslant k_c \leqslant 30$.
	Panel (c): pdfs of $\Pi$ at $k_c=8$, $k_c=20$ and $k_c=70$ during stage (II).}}
  \label{fig:TOT}
\end{figure*}

Since $\avg{\Pi}_{V,N} \geqslant 0$, it can be expected that the pdf of
$\Pi$ is positively skewed such that events leading to a forward
transfer of total energy across the filter scale are more likely than
backscatter events. This is indeed the case as shown by the \ml{standardized} pdf of $\Pi$ 
in Fig.~\ref{fig:TOT}(a) at $k_c=20$ for stages (I)-(III).  
Apart from more pronounced tails occurring in stage (II),
the \ml{standardized} pdfs are remarkably similar.  
{However,} while the pdf of $\Pi$ is positive skewed 
at all stages, the pdf of $\Pi^L$ is symmetric 
(see Fig.~\ref{fig:TOT_LEO} in Appendix 
). Hence, by measuring  
$\Pi = \Pi + \Pi^L$ as the total SGS energy transfer, the 
residual transfer amongst the resolved scales carried by the Leonard component 
could lead to the conclusion 
{of backscatter events being more frequent}
than they actually are. 

\mb{In Fig.\ref{fig:TOT}(b) we show the flatness of $\Pi$, ($F_{_\Pi}$), as a
function of the cut-off wavenumber $k_c$. From this analysis we can see that
the flatness
shows a similar power-law behavior
\ml{in the inertial range $2\le k_c\le 30$}
during all stages in the evolution. The
flatness scaling exponent $\zeta = 0.55 \pm 0.05, $ see inset of
Fig.~\ref{fig:TOT}(b), has been measured by a least-squares fit and its error has
been estimated by \ml{varying the fitting interval within the inertial range $2 \le k_c \le 30$.}
A small temporal variability of the flatness is observed only in the dissipative range where it is also found to increase exponentially suggesting strong deviation from Gaussianity at all
times. From both the pdfs and flatness analysis it follows that the statistical
properties of $\Pi$ are conserved during the temporal evolution. \ml{Fig.~\ref{fig:TOT}(c) presents} the pdfs of $\Pi$ at a fixed time during the non-linear stage (II) at three different cut-off wavenumbers \ml{$k_c=8$, $k_c=20$ and $k_c=70$. As can be seen from Figs.~\ref{fig:En_evo}(b) and (\ref{fig:TOT_mean}), $k_c=8$ and $k_c=20$ correspond to the beginning 
and the end of the inertial range, respectively, while $k_c=70$ lies in the dissipative range.}
From the comparison of the three standardized pdfs in Fig.\ref{fig:TOT}(c) we can
clearly observe the presence of intermittency in the statistics of $\Pi$ \ml{through an increasing} departure from Gaussianity at successively smaller scales.}
\mb{The same information can be extracted by the power-law behavior of the flatness over the inertial range of scales which also shows the intermittent properties of the SGS energy transfer in MHD turbulence. It is interesting to note that the value of $\zeta$ measured from the data is in agreement with the prediction of the She-Leveque model \cite{she1994universal}. Indeed, from the scaling estimate, 
$$\langle |\Pi|^n \rangle = O\left( \Delta^{\zeta_{3n}-n}\right),$$
\cite{eyink1996multifractal,Buzzicotti18a} and from the She-Leveque values of the exponents for n=2: $\zeta_6 \sim 1.77$ and n=4: $\zeta_{12} \sim 1.94$ (note that there is a typo in the value of $\zeta_{12}$ reported in ref.~\cite{Buzzicotti18a}), we obtain for the flatness the She-Leveque prediction $\zeta_{_{SL}} \sim 0.6$.}


\subsection{Kinetic and magnetic SGS energy transfers}

\begin{figure}[htbp]
  \centering
  \includegraphics[width=0.5\textwidth]{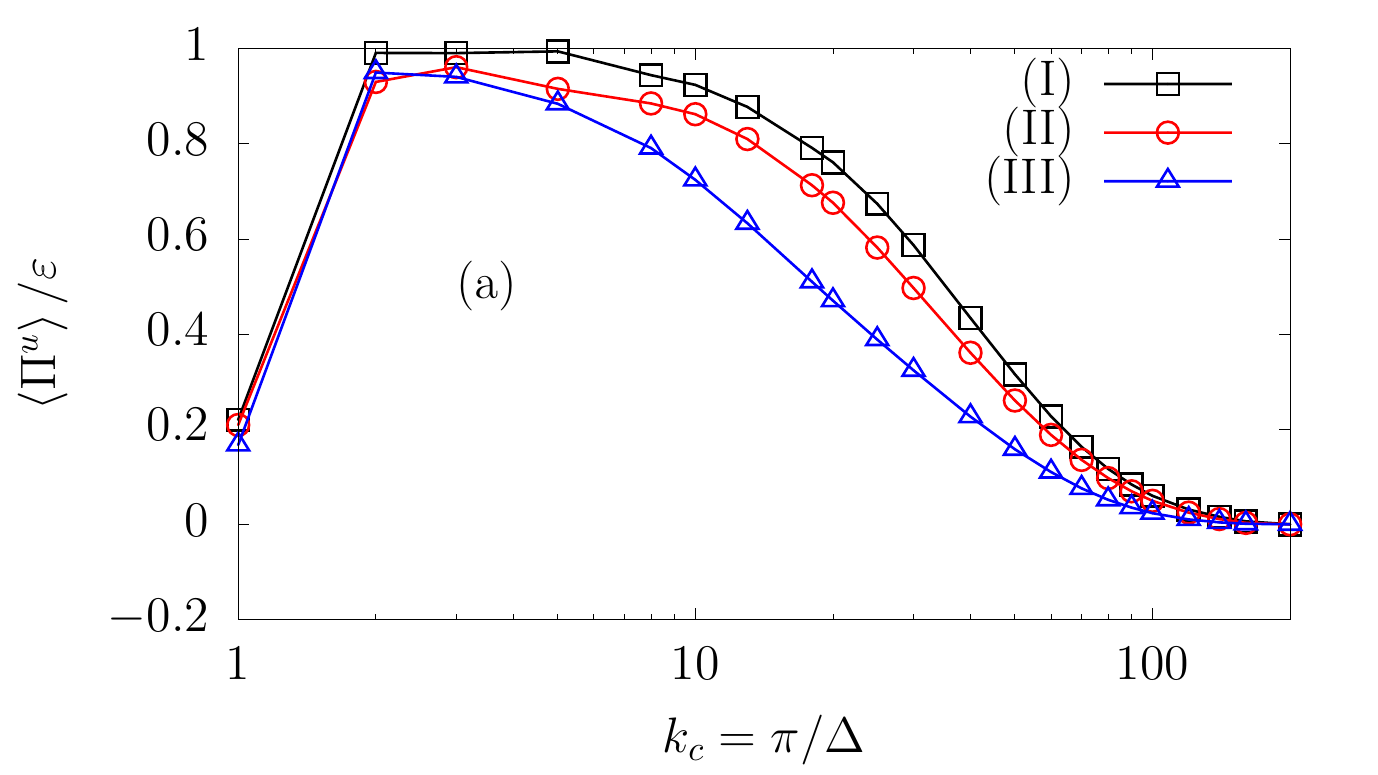}
  \includegraphics[width=0.5\textwidth]{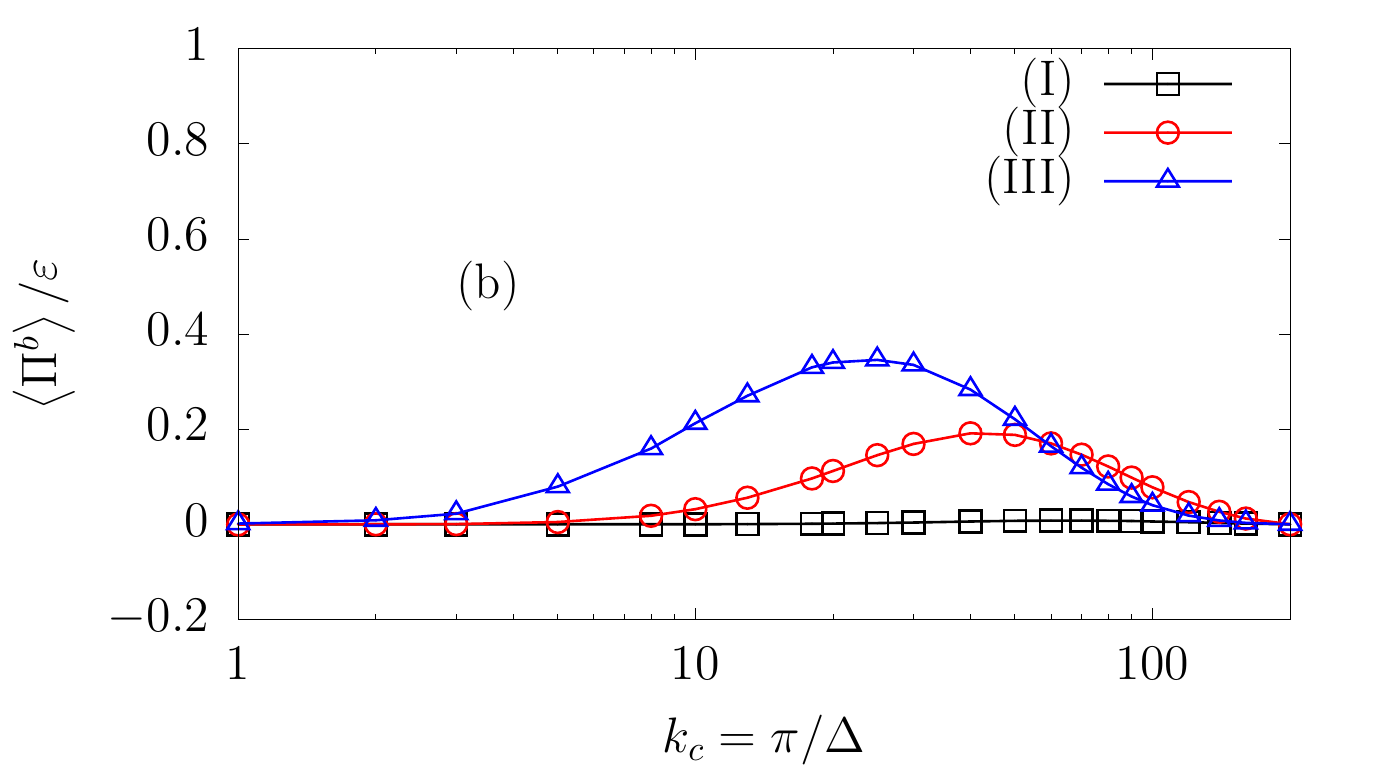}
  \includegraphics[width=0.5\textwidth]{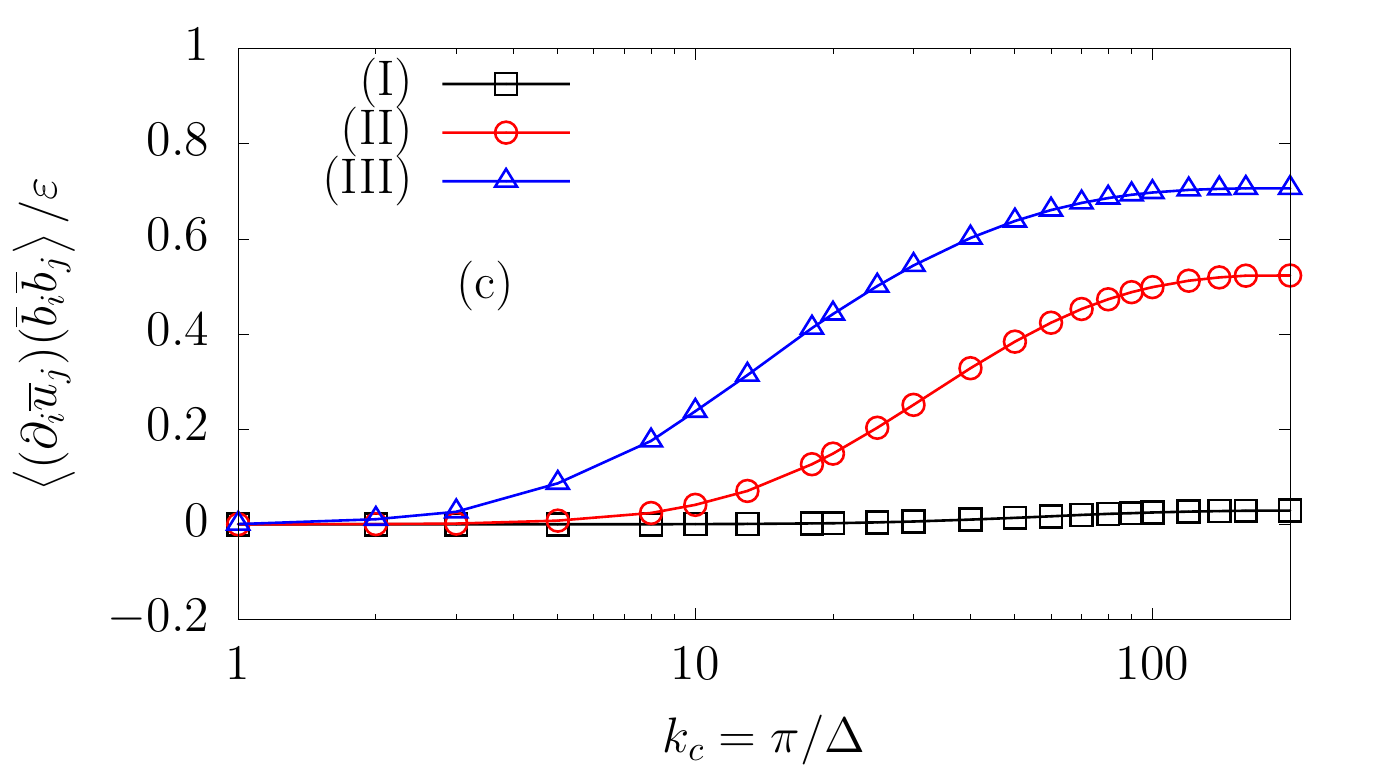}
  \caption{
	The mean 
    P-SGS energy transfers $\langle \Pi^{u}\rangle_{V,N}$ (panel (a)),
    and $\langle \Pi^{b}\rangle_{V,N}$ (panel (b)), and the mean
	of resolved-scale conversion term
	$\langle (\p{j}\overline{u_i})\overline{b}_i\overline{b}_j\rangle_{V,N}$ (panel (c)), 
    normalized with the total energy dissipation rate $\varepsilon$ 
    versus the cutoff wavenumber $k_c$ during the kinematic
	(I), non-linear (II) and stationary (III) stages.}
  \label{fig:VTOTBTOT_mean}
\end{figure}

As discussed in Sec.~\ref{sec:theory}, $\Pi$ can be further decomposed
into 
$\Pi^{u}$ and 
$\Pi^{b}$.  Furthermore, the resolved-scale conversion term,
$(\p{j}\ou_i)\ob_i\ob_j$, {in Eqs.~\eqref{eq:evol_Eu} and \eqref{eq:evol_Eb}}, 
{which cancels out in Eq.~\eqref{eq:total_ples} for the total resolved-scale energy, 
must also be measured.}
{It contains information on the scale-dependence of the conversion of kinetic 
to magnetic energy, an assessment of which is essential in order to provide guidance
for SGS models of MHD dynamos.}   
\\
The averages $\avg{\Pi^{u}}_{V,N}$, $\avg{\Pi^{b}}_{V,N}$ and
$\langle (\p{j}\ou_i)\ob_i\ob_j\rangle_{V,N}$ are shown in Fig.~\ref{fig:VTOTBTOT_mean}(a-c),
respectively.  We first notice that $\avg{\Pi^{u}}_{V,N}$ gets depleted towards
stage (III) while $\avg{\Pi^{b}}_{V,N}$ increases.  From a comparison of the large
increase of {$\langle (\p{j}\ou_i)\ob_i\ob_j\rangle_{V,N}$} 
relative to the smaller decrease of
$\avg{\Pi^{u}}_{V,N}$ during stages (I)-(III), it follows that the growth of the
magnetic field is due to direct interactions between $\overline{\vec{u}}$ and
$\overline{\vec{b}}$.  
\ml{The data presented in Fig.~\ref{fig:VTOTBTOT_mean}(a,b)}
also show that both the kinetic and magnetic
SGS energy transfers are forward. 
\ml{From Fig.~\ref{fig:VTOTBTOT_mean}(c) it can be seen}
that 
\blue{ $\langle (\p{j}\ou_i)\ob_i\ob_j\rangle_{V,N}$ 
has an inflection point that saturates at $k^* \approx 20$. Since the 
large-scale conversion term is the running integral in $k$ 
of the energy transfer at $k$, 
an inflection point in $\langle (\p{j}\ou_i)\ob_i\ob_j\rangle_{V,N}$ at $k^*$ 
implies an extremum in the energy conversion at $k^*$, corresponding 
to a saturation length scale for the conversion of kinetic to magnetic energy.}
The existence of a
saturation length scale implies the breaking of inertial self-similarity and
puts a natural constraint on any LES for MHD. Either we use an extremely
resolved model with $k_c\gg k^*$, and we fully resolve the dynamics leading to
the non-linear dynamo saturation, or we use $k_c \sim k^*$ and a very
sophisticated SGS model must be used. Certainly one cannot further push and use
$k_c \ll k^*$, or a fully ad-hoc magnetic field growth must be supplied.  
\ml{An in-depth investigation of the statistical properties of 
    the resolved-scale conversion term would provide guidance 
    for cases where very coarse grids require the aforementioned 
    ad-hoc magnetic forcing term. 
}
A quantitative assessment of this issue \ml{also} 
requires {\em a posteriori} analyses and would 
constitute a useful contribution to MHD LES.

\begin{figure}[htbp]
  \centering
  \includegraphics[width=0.5\textwidth]{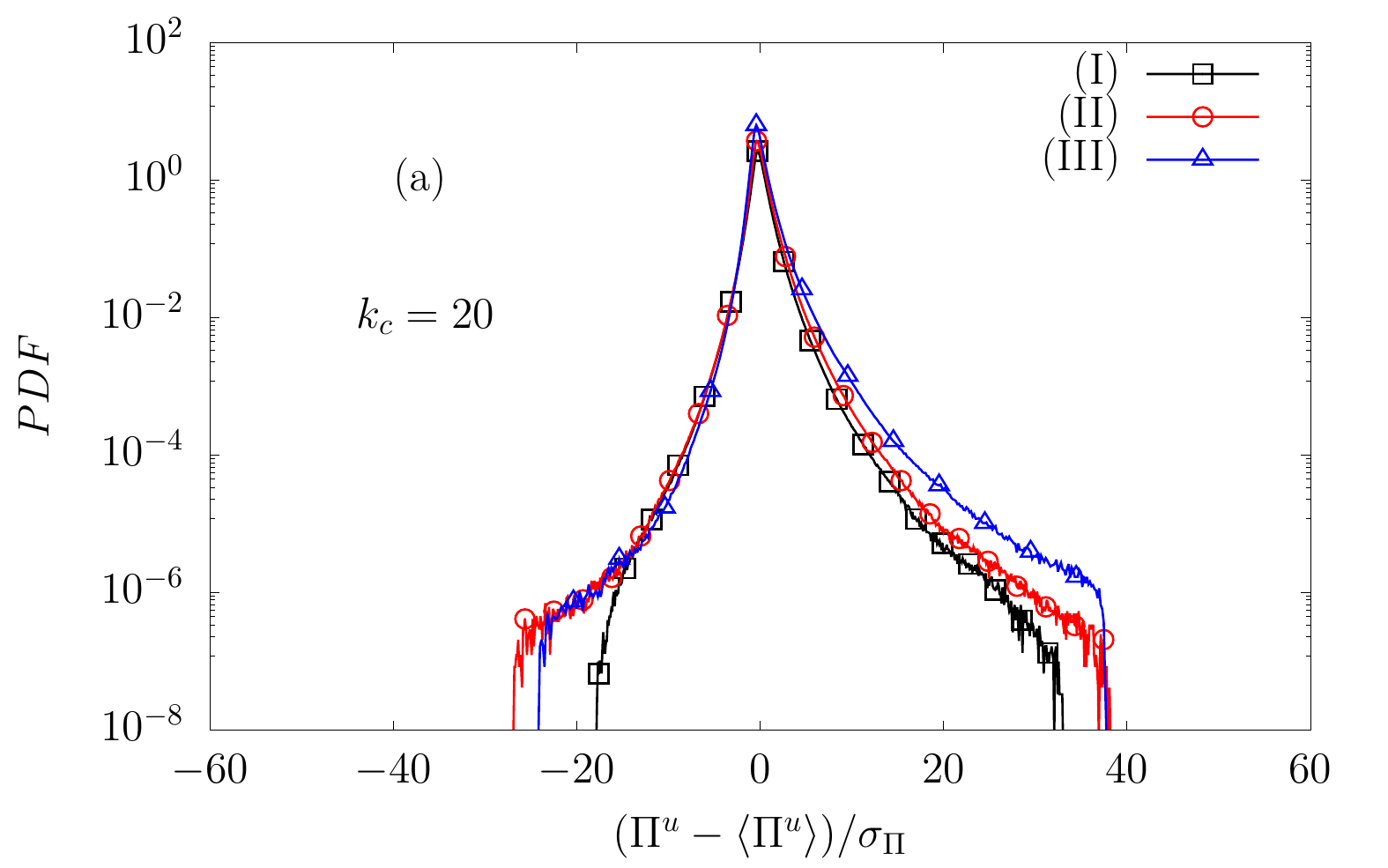}
  \includegraphics[width=0.5\textwidth]{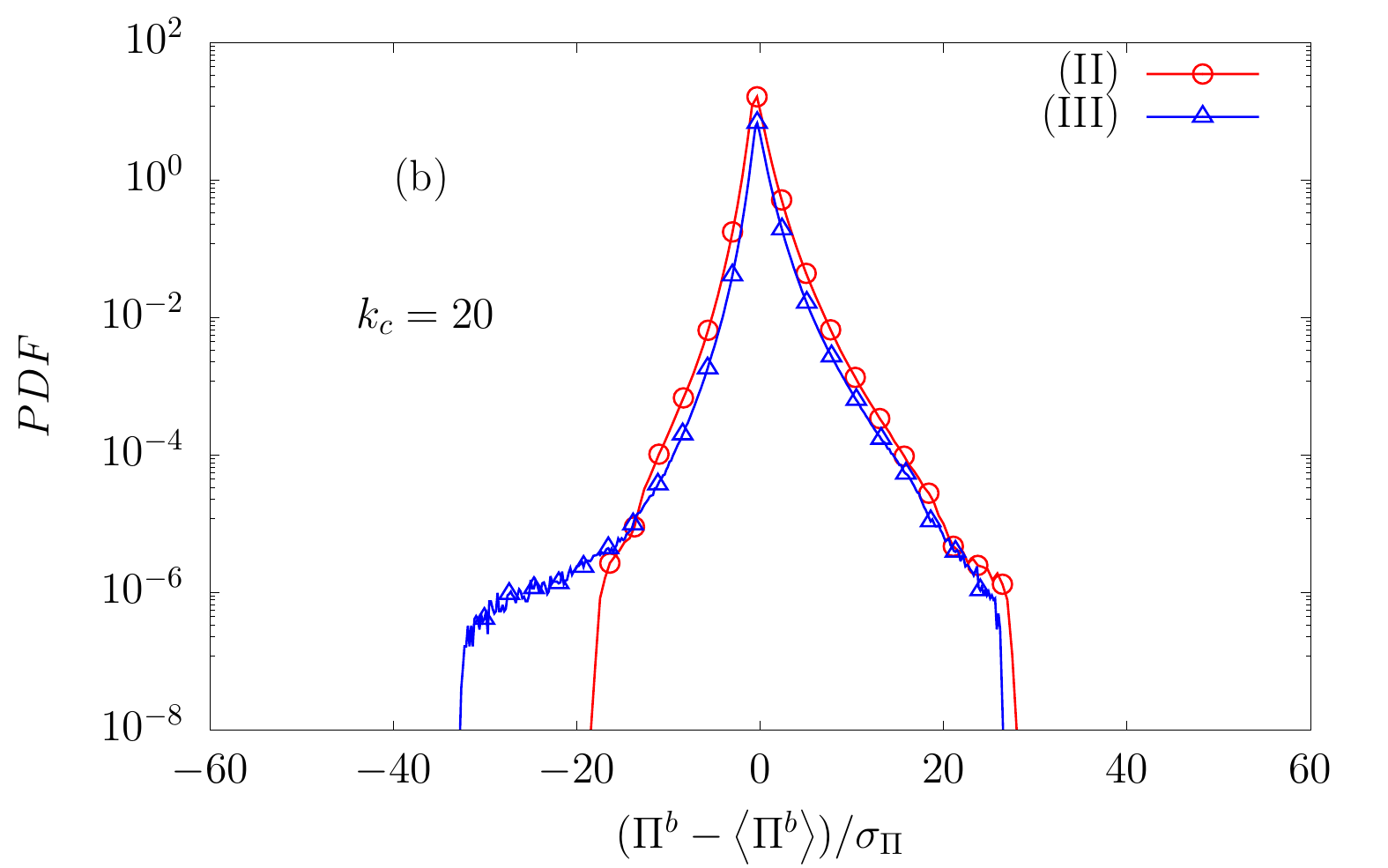}
  \includegraphics[width=0.5\textwidth]{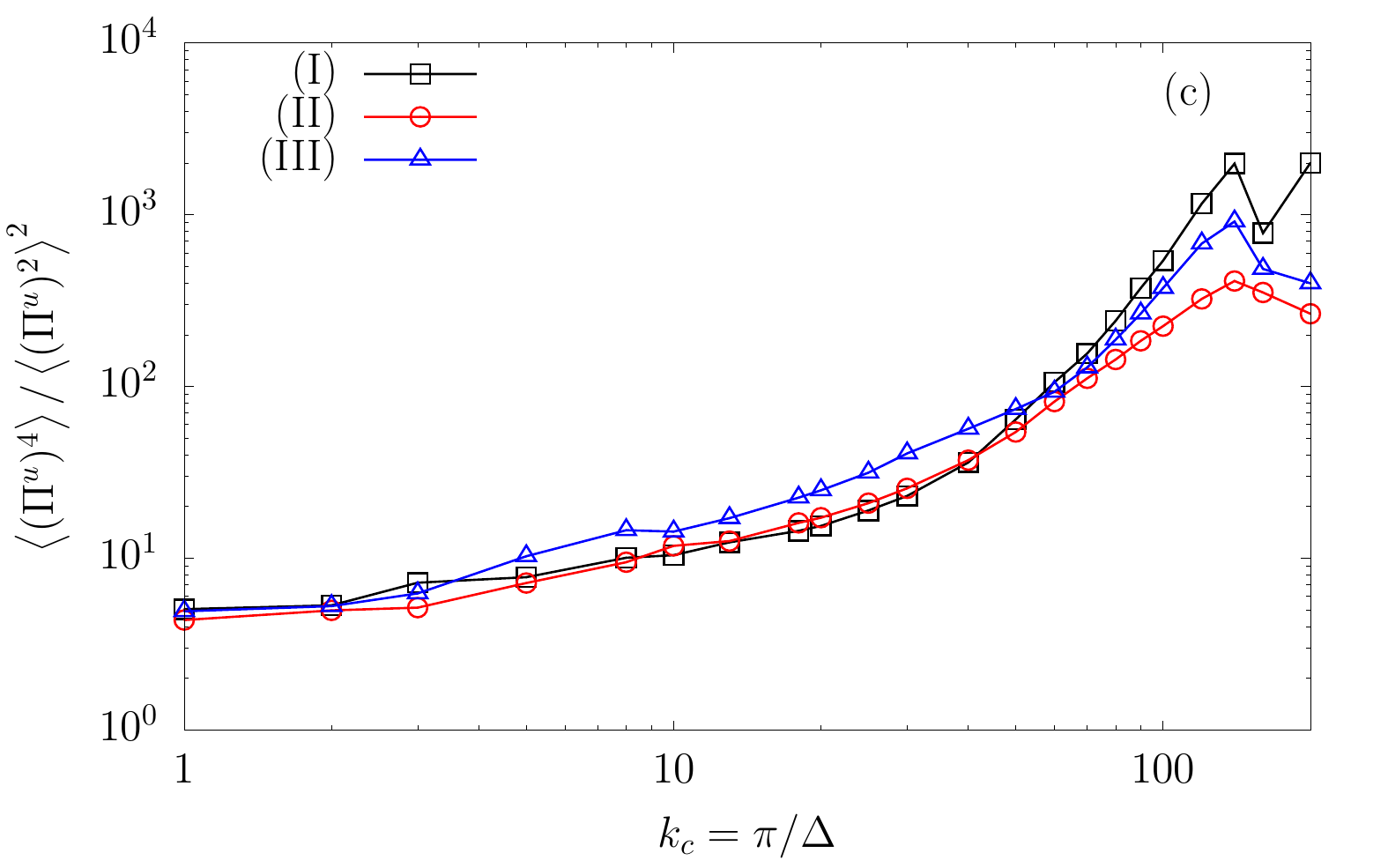}
  \includegraphics[width=0.5\textwidth]{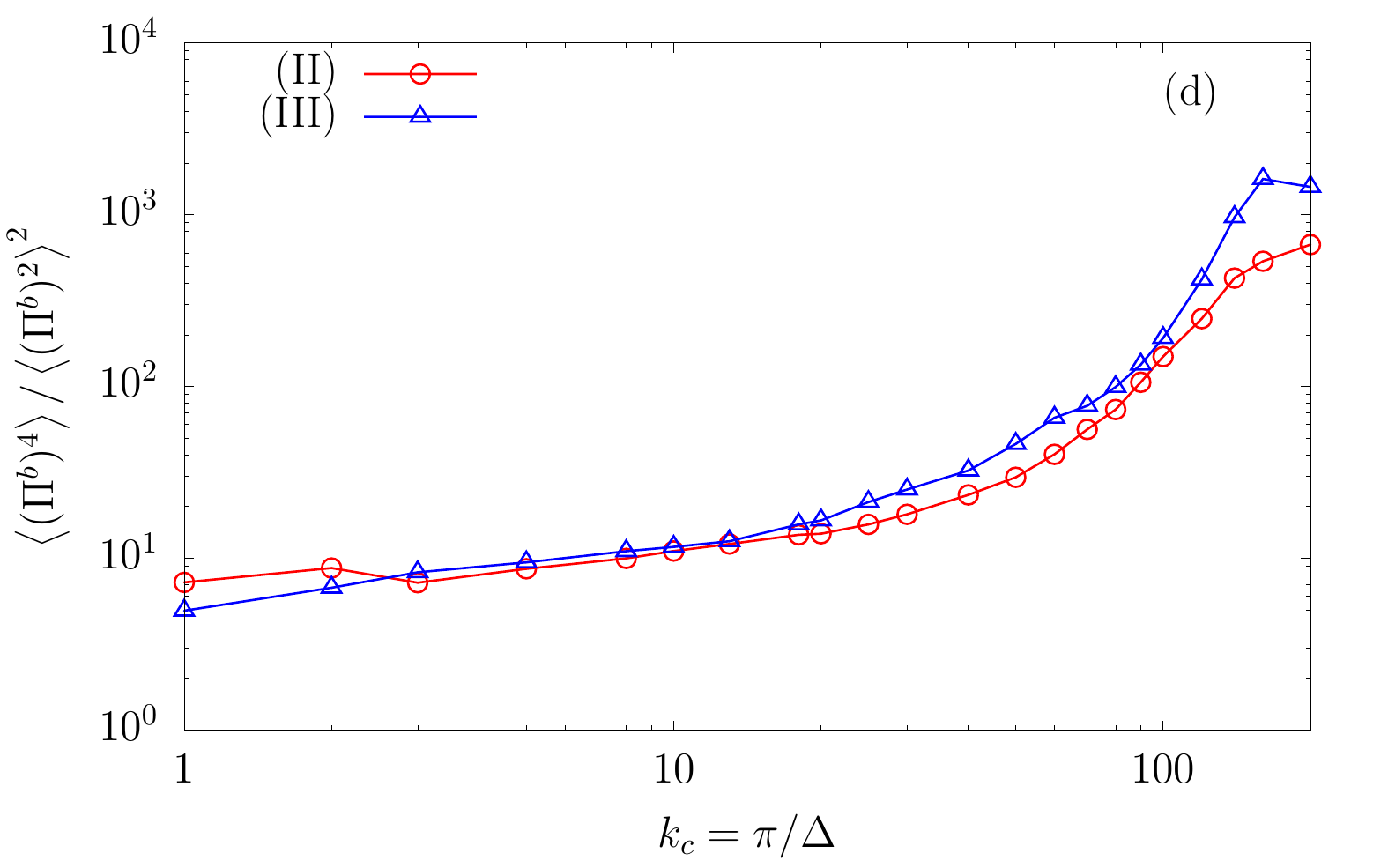}
  \caption{
        Kinetic and magnetic SGS energy transfers $\Pi^{u}$ and $\Pi^{b}$ during the
        kinematic stage (I), non-linear stage (II) and the stationary stage (III):
	pdfs of $\Pi^{u}$ (a) and $\Pi^{b}$ (b) at $k_c=20$;
	flatness of $\Pi^{u}$ (c) and $\Pi^{b}$ (d) against the
	cutoff wavenumber $k_c$.}
  \label{fig:VTOTBTOT}
\end{figure}

Figure~\ref{fig:VTOTBTOT}(a,b) presents the standardized pdfs of
$\Pi^{u}$ and $\Pi^{b}$ at $k_c=20$. {We note that the pdfs of $\Pi^{b}$ are only
shown for stages (II) and (III), as $\Pi^{b}$ is negligible in stage  (I), because the system
is dominated by magnetic field amplification which occurs through the term $(\p{j}\overline{u_i})\overline{b}_i\overline{b}_j$.}
{Although $\langle \Pi^{u}\rangle_{V,N}$ and $\langle \Pi^{b}\rangle_{V,N}$
are positive, that is, kinetic and magnetic energies are transferred downscale on average, 
the pdfs of $\Pi^{u}$ and $\Pi^{b}$ develop negative tails. The latter is particularly pronounced 
for $\Pi^{b}$ in stage (III), as shown in Fig.~\ref{fig:VTOTBTOT}(b).}
{That is} backscatter events in the 
magnetic SGS energy transfer cannot be neglected {for a fully nonlinear dynamo}. The latter implies that 
dissipative {approaches} 
such as the Smagorinsky {closure} \cite{Smagorinsky63} are 
hardly optimal to model the SGS stresses in the induction equation.
The flatness of $\Pi^{u}$ and $\Pi^{b}$ as a function of $k_c$ is shown 
Figs.~\ref{fig:VTOTBTOT}(c,d). 
There {appears to be} a slight indication of increased  
intermittency in stage (III) compared to stages (I) and (II) for both $\Pi^{u}$ and $\Pi^{b}$
{since the flatness becomes more scale-dependent in the inertial range. 
As can be seen from the 
figures, $\Pi^b$ appears to be less intermittent than $\Pi^u$.
However, the latter statements on intermittency require further assessment 
using higher-resolved datasets with a more extended inertial range}. 

\begin{figure*}[htbp]
 \centering
 \includegraphics[width=0.9\textwidth]{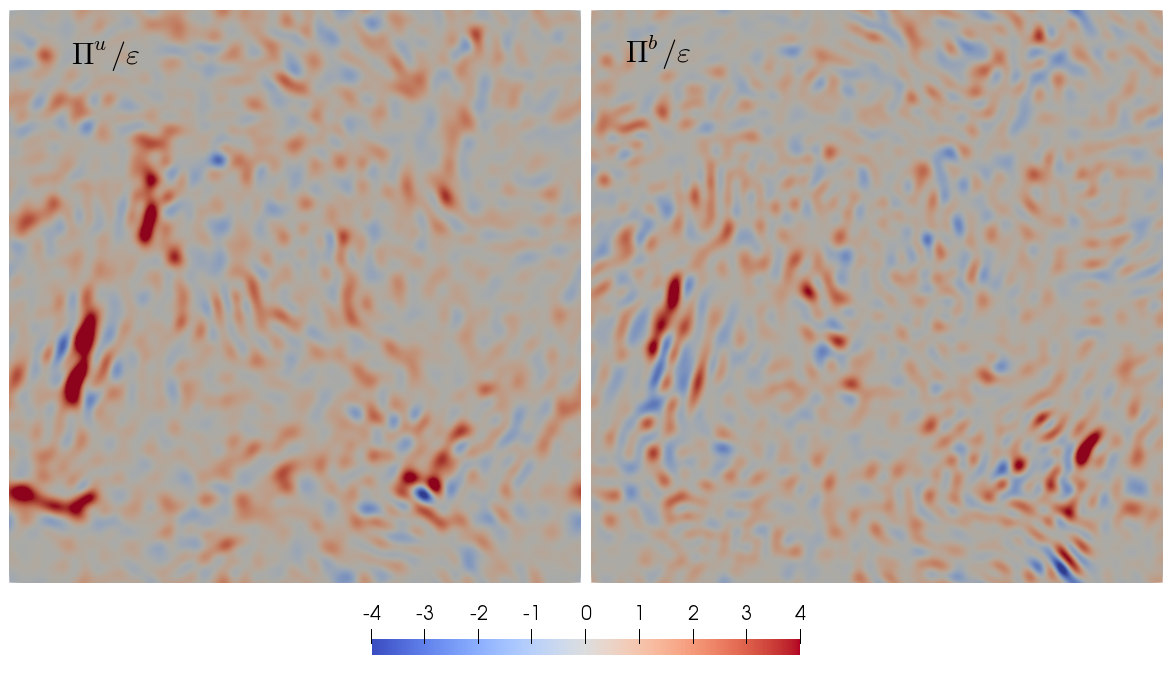}
 \includegraphics[width=0.9\textwidth]{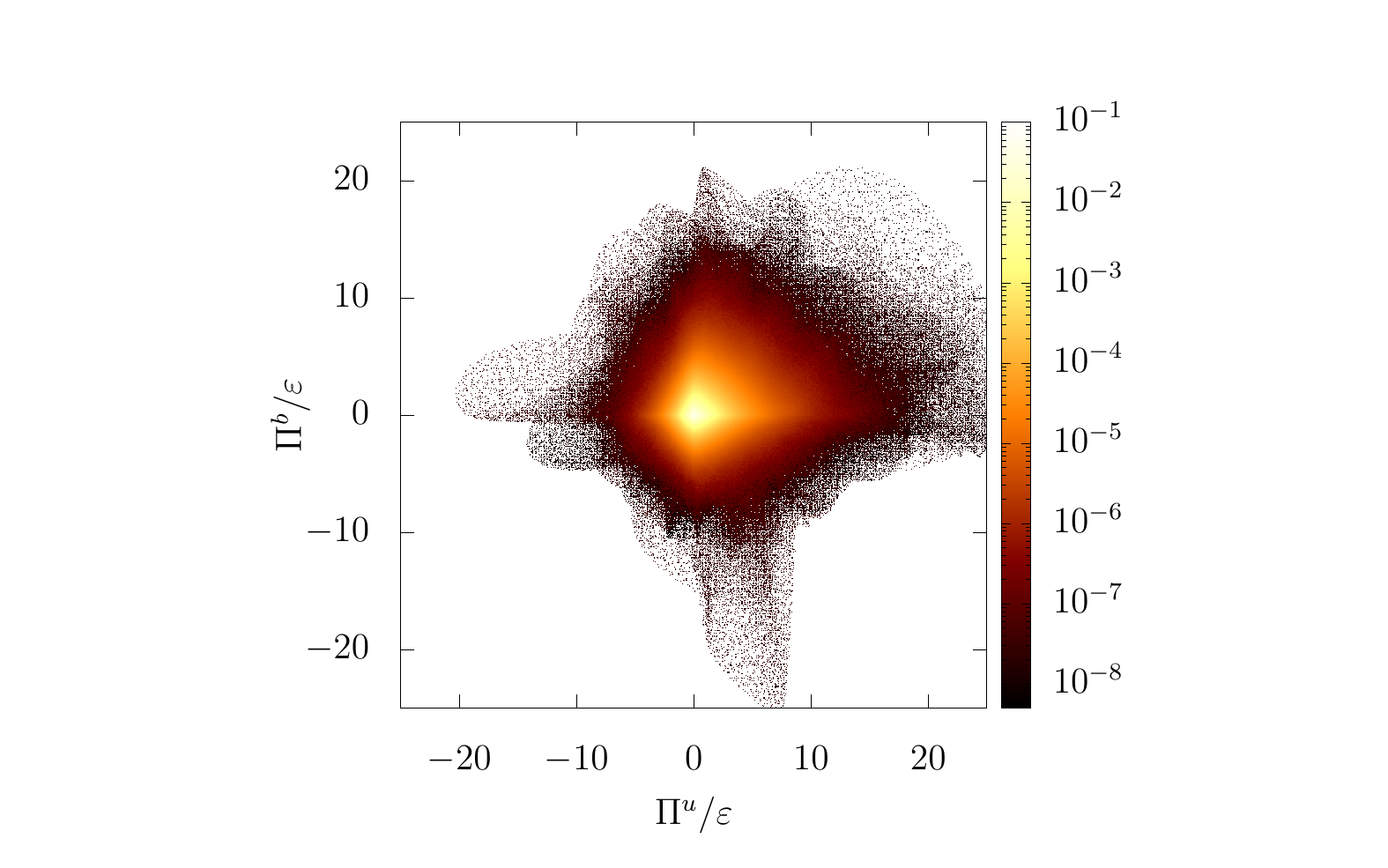}
 \caption{(Colour online)
   {Top:} Two-dimensional visualisations of 
   $\Pi^{u}$ (left) and $\Pi^{b}$ (right)  at $k_c=20$ 
   in stage (III). 
   Positive values correspond to forward energy transfer while negative values
   indicate backscatter. 
   {Bottom: Corresponding joint pdf of $\Pi^{u}$ and $\Pi^{b}$.}
 }
 \label{fig:Visu_Piub}
\end{figure*}

Visualisations of $\Pi^{u}$ and $\Pi^{b}$ obtained during stage (III) 
are presented in {the top panels of} Fig.~\ref{fig:Visu_Piub}. 
A striking feature is the localized \ml{elongated} 
nature 
of intense forward-transfer events in $\Pi^{u}$. Similar 
structures are 
are also visible in $\Pi^{b}$
, and the colour-mapping suggests an \blue{inverse relation} 
between $\Pi^{u}$ and $\Pi^{b}$, \blue{where large values of $\Pi^{u}$ are correlated with small values of $\Pi^{b}$
and vice versa}.
The correlation between $\Pi^{u}$ and $\Pi^{b}$ is quantified
through their joint pdf shown in the bottom panel of Fig.~\ref{fig:Visu_Piub}. 
\ml{The data in the figure show a tendency towards higher probabilities along the axes where either 
$\Pi^{u} = 0$ or $\Pi^{b} = 0$, which suggest a mild \blue{inverse proportionality} 
between the two. 
}
As will be seen later, the intense forward-transfer events in $\Pi^{u}$ originate
from the {P-SGS} Maxwell stresses in 
{Eq.~\eqref{eq:momentum_PLES}}.

\subsection{Inertial and Maxwell SGS energy transfers}
\begin{figure}[htbp]
  \centering
  \includegraphics[width=0.5\textwidth]{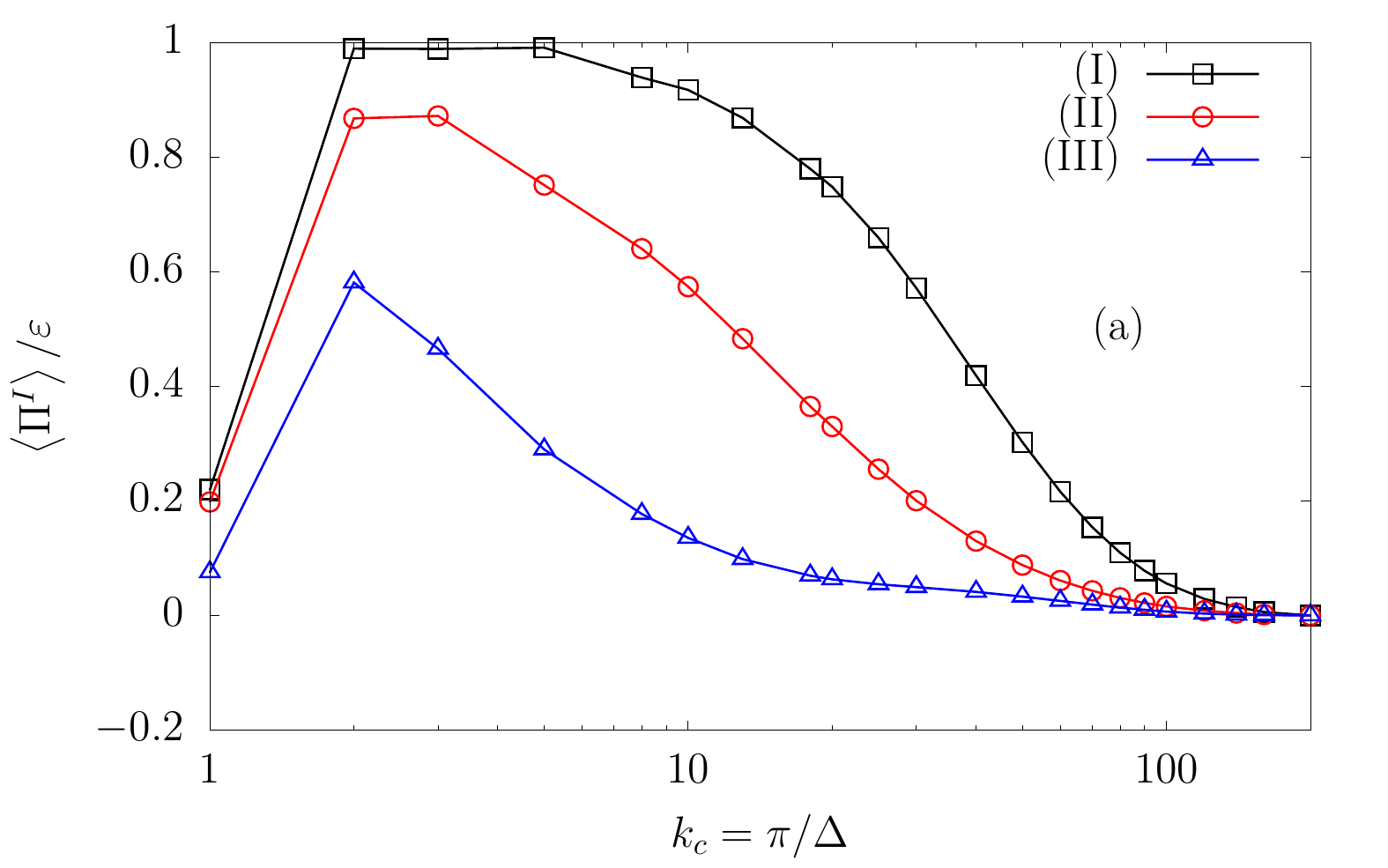}
  \includegraphics[width=0.5\textwidth]{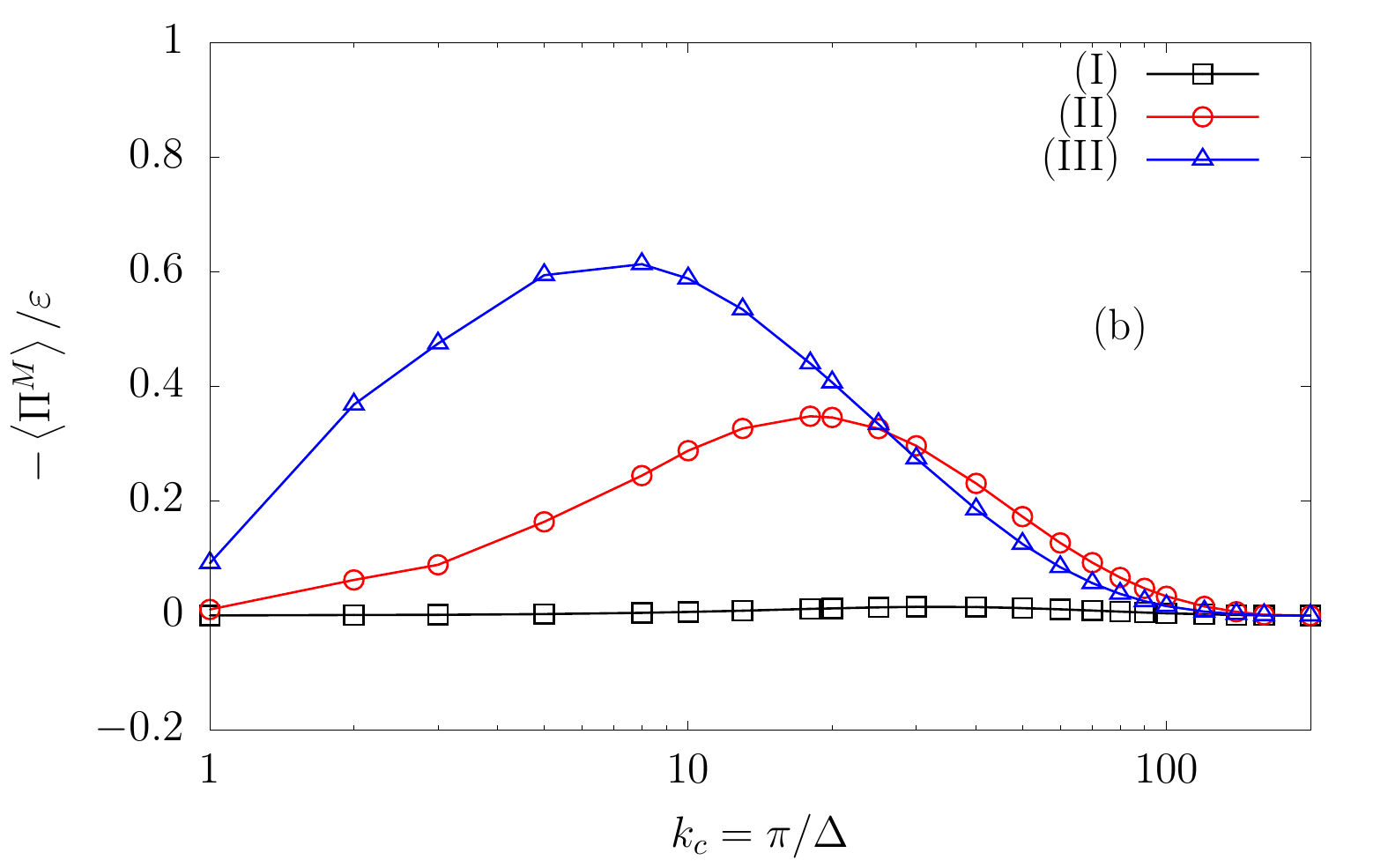}
  \caption{
	The mean components 
    $\avg{\Pi^{I}}_{V,N}$ (a)
	and 
    $\avg{\Pi^{M}}_{V,N}$ (b) versus the cutoff
	wavenumber $k_c$ at the kinematic stage (I), non-linear stage (II) and the
	stationary stage (III).}
  \label{fig:VELMAG_mean}
\end{figure}

The term {$\Pi^{u}$ in Eq.~\eqref{eq:momentum_PLES}} 
is now further decomposed into 
$\Pi^{I}$ and 
$\Pi^{M}$, as introduced in Sec.~\ref{sec:theory}. 
Figure~\ref{fig:VELMAG_mean} presents $\avg{\Pi^{I}}_{V,N}$ and $\avg{-\Pi^{M}}_{V,N}$
as functions of $k_c$ {where the sign convention for $\Pi^{M}$ reflects
the sign with which it occurs in Eq.~\eqref{eq:momentum_PLES}}. During the kinematic stage (I), $\avg{\Pi^{M}}_{V,N}$ is 
negligible and the total SGS energy transfer is carried by $\avg{\Pi^{I}}_{V,N}$.  
As expected $\langle \Pi^{I}\rangle_{V,N}$ gets depleted towards stage (III) while 
$\langle -\Pi^{M} \rangle_{V,N}$ increases. 
Both $\avg{\Pi^{I}}_{V,N}$ and $\avg{-\Pi^{M}}_{V,N}$ are positive, that is, 
the resolved-scale kinetic energy is transferred from large to small scales through inertial transfer 
as well as through the Maxwell component.


\begin{figure}[htbp]
  \centering
  \includegraphics[width=0.5\textwidth]{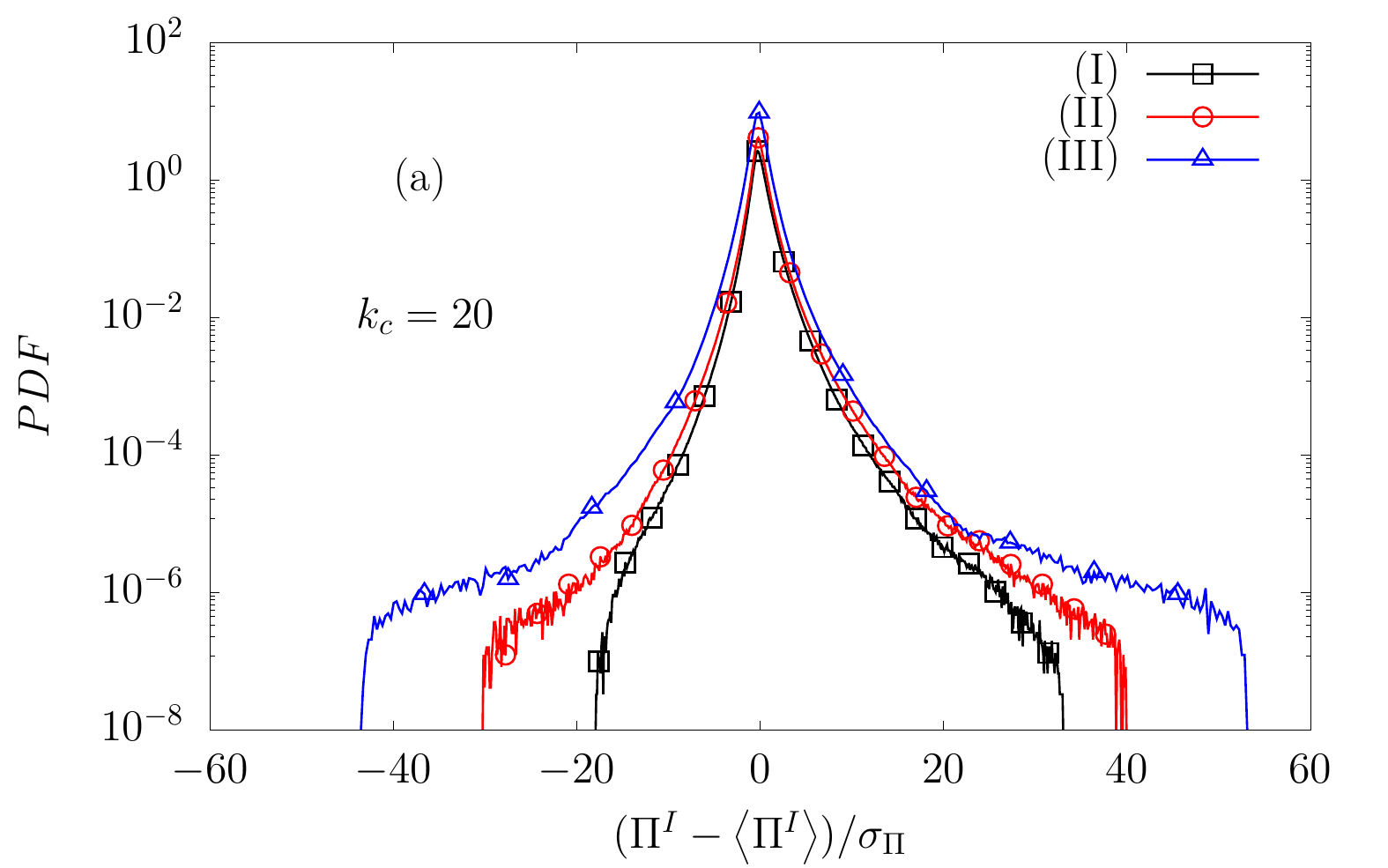}
  \includegraphics[width=0.5\textwidth]{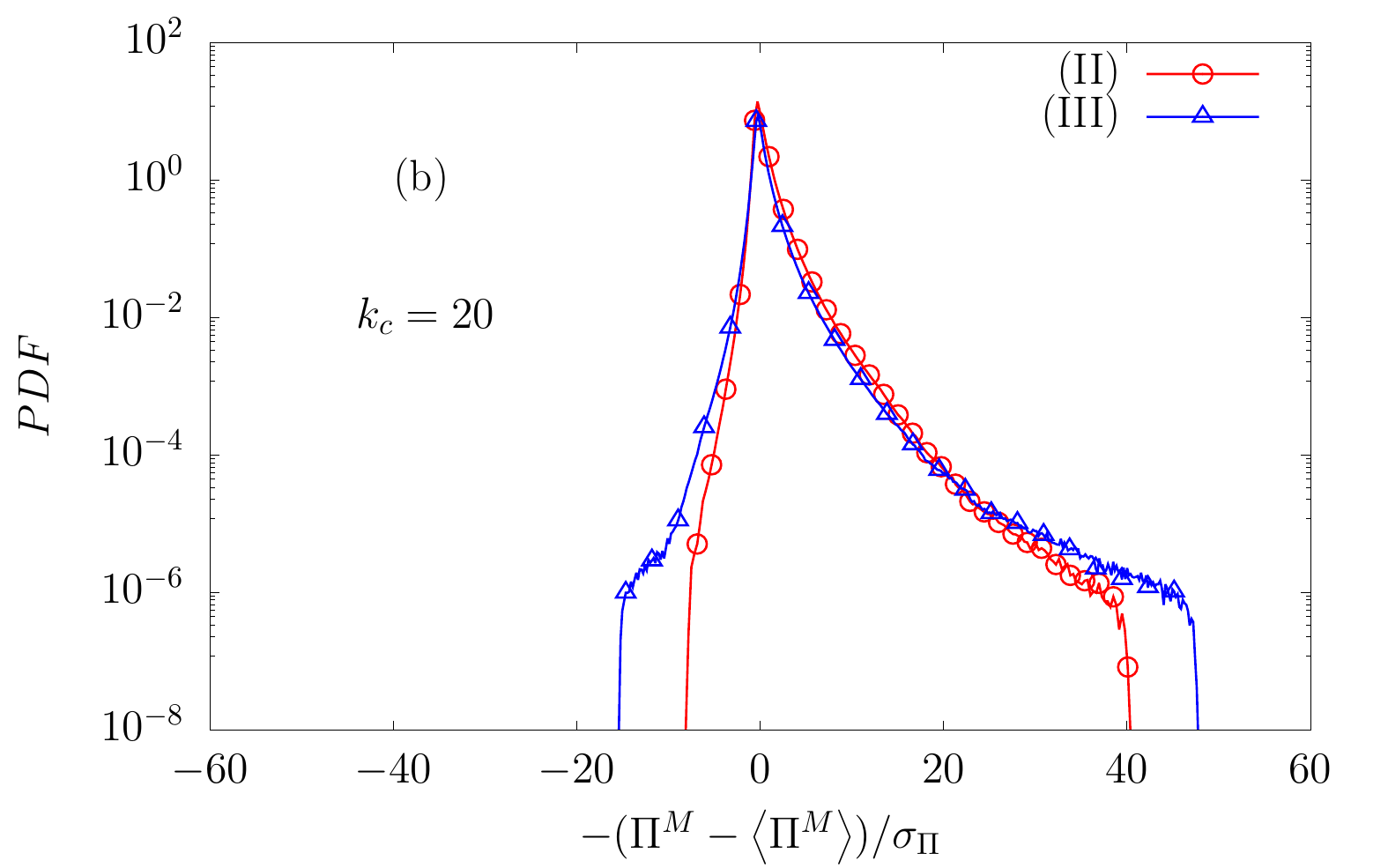}
  \includegraphics[width=0.5\textwidth]{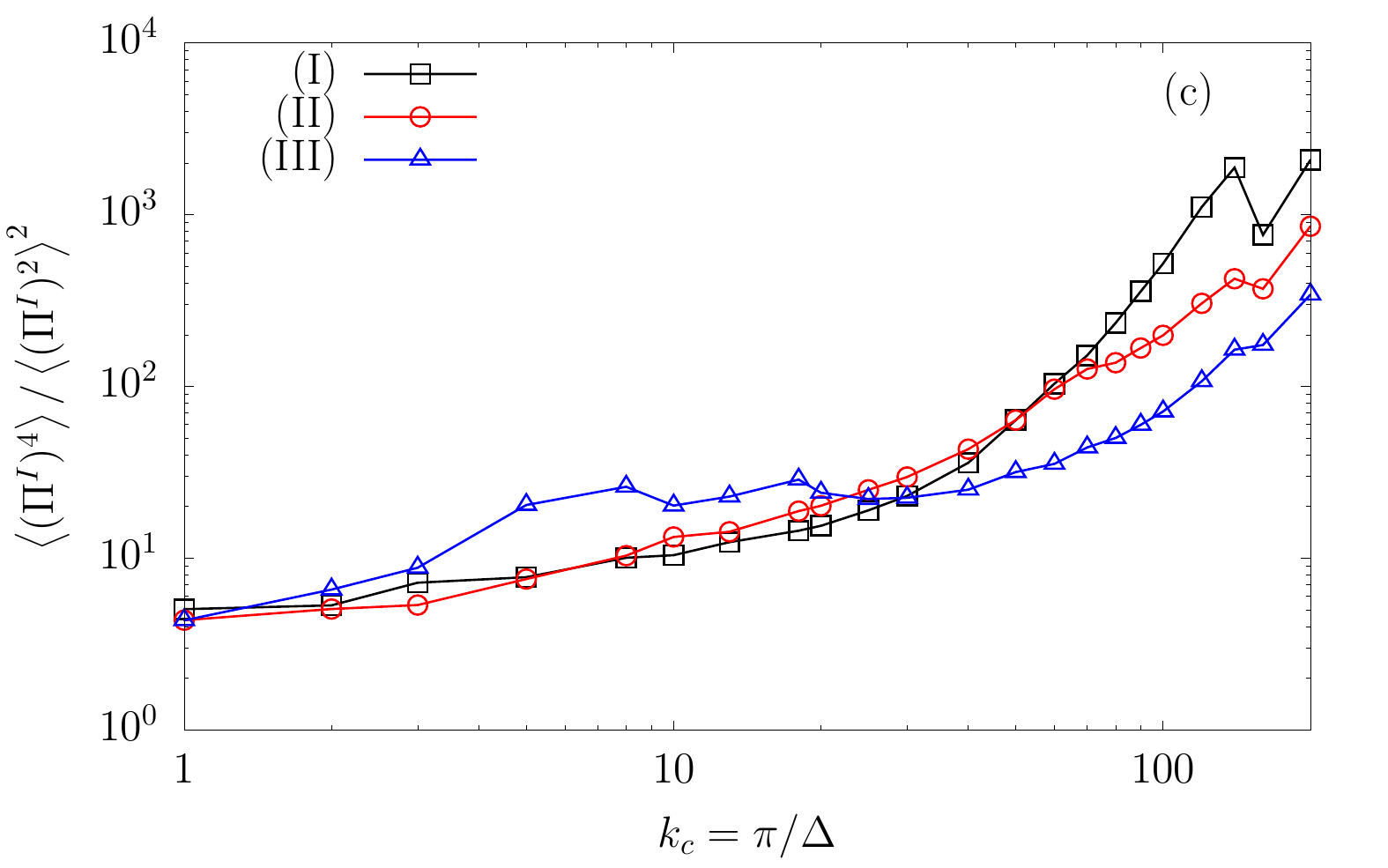}
  \includegraphics[width=0.5\textwidth]{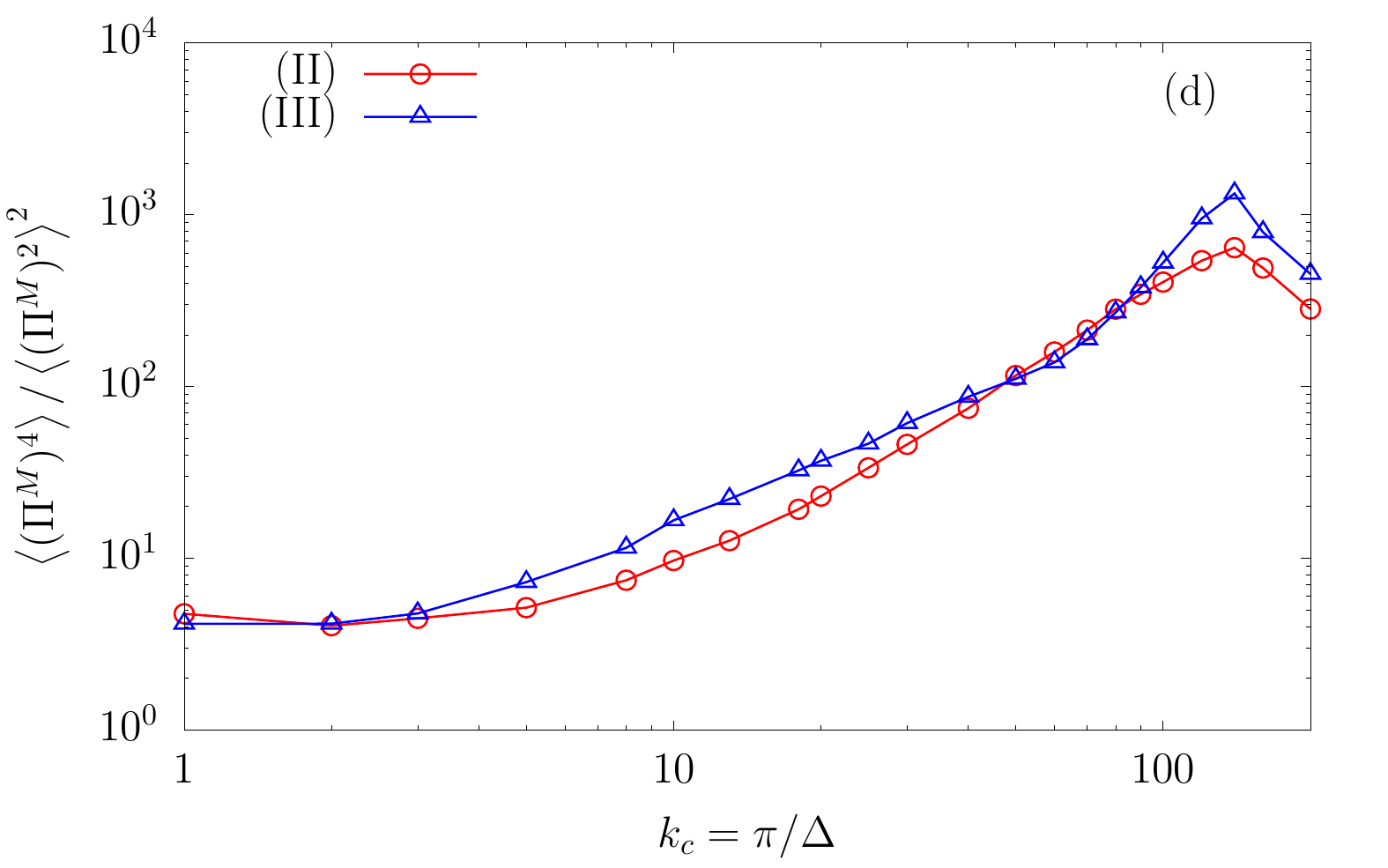}
  \caption{
        {Fluctuations of}
        $\Pi^{I}$ and $\Pi^{M}$ during
        kinematic (I), non-linear (II) and stationary (III) stages:
	pdfs of $\Pi^{I}$ (a) and $\Pi^{M}$ (b) at $k_c=20$;
	flatness of $\Pi^{I}$ (c) and $\Pi^{M}$ (d) against the
	cutoff wavenumber $k_c$.
        }
  \label{fig:VELMAG}
\end{figure}

Figures~\ref{fig:VELMAG} (a,b) show the standardized pdfs of 
{$\Pi^{I}$ and $\Pi^{M}$, respectively, where we note that 
the pdf of $\Pi^{M}$ is only shown in stages (II) and (III) as it 
is negligible in stage (I)}. 
During stages (II) and (III) the pdf of $\Pi^{I}$ changes significantly
compared to its shape during stage (I), where the inertial dynamics
are approximately unaffected by the magnetic field. The most striking 
feature here is the development of wide tails and a much more symmetric
shape. That is, the inertial SGS energy transfer fluctuates very differently
in presence of a fluctuating magnetic field as in the nonconducting case: 
First, the wide tails indicate that extreme events are more likely than in 
the nonconducting case. Second, the symmetric shape implies that backscatter
events in the inertial SGS energy transfer become significant. 
%
In contrast, as can be seen from Fig.~\ref{fig:VELMAG} (b), 
the pdf of $\Pi^{M}$ has a clear positive skewness. That is, 
backscatter events are much less important than for all other
SGS energy transfer components and the contributions from the SGS 
Maxwell stresses should be well approximated by a dissipative model.

Measurements of the pdfs of ${\Pi^{I} + \Pi^{I,L}}$ and $\Pi^{M} + \Pi^{M,L}$ during the saturated
stage of a small-scale dynamo have been reported recently \cite{Kessar16}.  
By comparison of Figs.~\ref{fig:VELMAG} (a,b) with the left panel of Fig.~7 in 
Ref.~\cite{Kessar16}, one observes that the shape of the pdfs measured in Ref.~\cite{Kessar16}
is quite different from the results found here {for $\Pi^{I}$ and $\Pi^{M}$} . More precisely, 
the pdf of $\Pi^{I} + \Pi^{I,L}$ in Ref.~\cite{Kessar16} 
lacks the wide tails seen {for $\Pi^{I}$} here,
and the pdf of $\Pi^{M} + \Pi^{M,L}$ 
is much more symmetric than that presented 
{for $\Pi^{M}$} in Fig.~\ref{fig:VELMAG}(b). 
{There are two reasons for latter difference. First,  
the Leonard component is included in the measurement of the SGS
energy transfer in Ref.~\cite{Kessar16} while it is not included here. 
Second, the Reynolds numbers and filter widths also differ. 
In Ref.~\cite{Kessar16} the the pdfs were measured at $\Rl =75$ at a filter scale coresponding to 
$k_c=64$. For comparison, in our dataset $\Rl =211$, and the pdfs in Figs.~\ref{fig:VELMAG} (a,b) 
are measured at $k_c=20$. Even in our simulations, it can be seen from 
the energy spectra (Fig.~\ref{fig:En_evo}(b)) and the mean SGS energy transfer 
(Fig.~\ref{fig:TOT_mean}) that the dynamics at $k_c=64$ is significantly affected 
by viscous and Joule dissipation. This will be even more so for lower $\Rl$.
In order to provide a like-for-like comparison, we measured of the pdfs of 
$\Pi^M$, $\Pi^{M,L}$ and $\Pi^{M} + \Pi^{M,L}$ for $k_c = 80$, which for our data at $\Rl =211$ 
is comparable to $k_c=64$ for $\Rl =75$.
}
As can be seen in Fig.~\ref{fig:MAG_LEO} in the Appendix, 
{the pdf of $\Pi^{M,L}$ in the viscous range is sizeable and symmetric, such that}
the inclusion of {$\Pi^{M,L}$} 
in the measurement of the Maxwell SGS transfer
masks the distinctive positive skewness of its PDF. 


Figures~\ref{fig:VELMAG}(c,d) present the flatness of $\Pi^{I}$ and $\Pi^{M}$
as functions of $k_c$. For $\Pi^{I}$, the development of 
{strongly non-Gaussian statistics} 
is also reflected in the flatness, which has higher values in stage (III) compared 
to stages (I) and (II). Furthermore, the flatness has a much weaker scale-dependence 
during stage (III) as shown in Fig.~\ref{fig:VELMAG}(c). This indicates a depletion of intermittency 
of the velocity field in presence of a saturated dynamo. 
{Indeed, a} comparison of the $p^{\rm th}$-order 
scaling exponents $\zeta_p$ of the velocity-field structure functions 
for hydrodynamic turbulence \cite{Gotoh02} and for a saturated MHD dynamo \cite{Haugen04} 
reveals differences in $\zeta_p$ for $p \geqslant 5$. According to these results, 
the velocity field is less intermittent in presence of a saturated dynamo, as
observed here. 
Since $(\Pi^{I})^p$ is related to the $3p^{\rm th}$-order velocity-field 
structure function \cite{eyink2} the scaling properties of high-order structure functions determine
the behavior of the flatness of $\Pi^{I}$. Therefore differences concerning intermittency between MHD and 
hydrodynamic turbulence are more clearly visible in measurements of the flatness 
of $\Pi^{I}$ compared to direct measurements of $\zeta_p$. 
However, a quantitative assessment of the scaling properties of the flatness 
of $\Pi^{I}$ requires a further extended scaling range.    
In contrast to the results for $\Pi^{I}$, the flatness of $\Pi^{M}$ shown in 
Fig.~\ref{fig:VELMAG}(d) retains its scale-dependence after 
dynamo saturation. 
As can be seen from the figure, the flatness of $\Pi^{M}$ has a much stronger scale 
dependence compared to $\Pi$. 
\ml{The stronger intermittent signal in $\Pi^{M}$ may be related to the 
fact that the saturated magnetic field is much more intermittent than the velocity 
field that maintains it, as shown by measurements of scaling exponents of inertial and magnetic
structure functions obtained from DNSs of stationary small-scale dynamos \cite{Haugen04}.}
As in the present data, no mean magnetic field was present in the data analysed in 
Ref.~\cite{Haugen04}. 

\begin{figure}[htbp]
  \centering
  \includegraphics[width=0.5\textwidth]{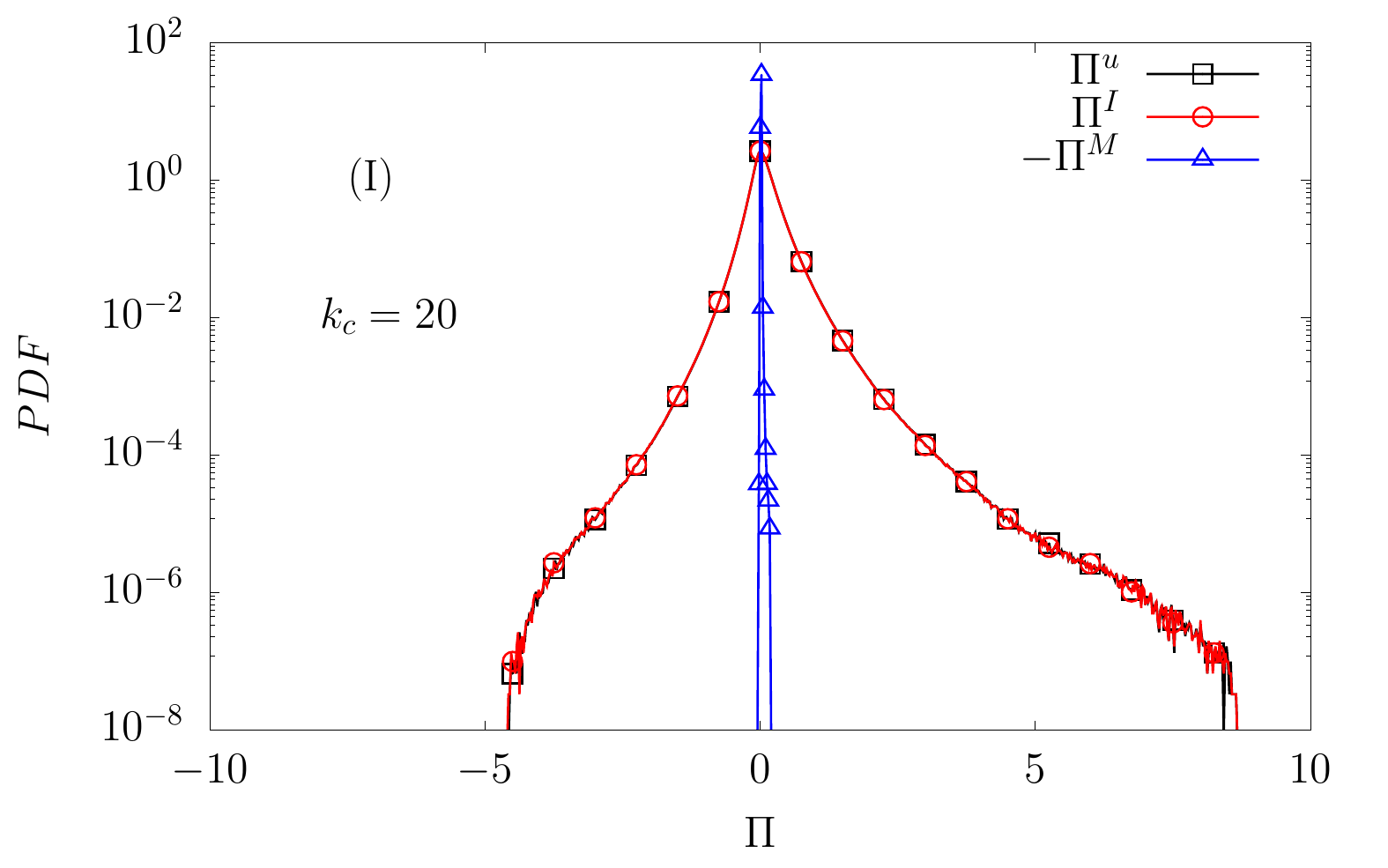}
  \includegraphics[width=0.5\textwidth]{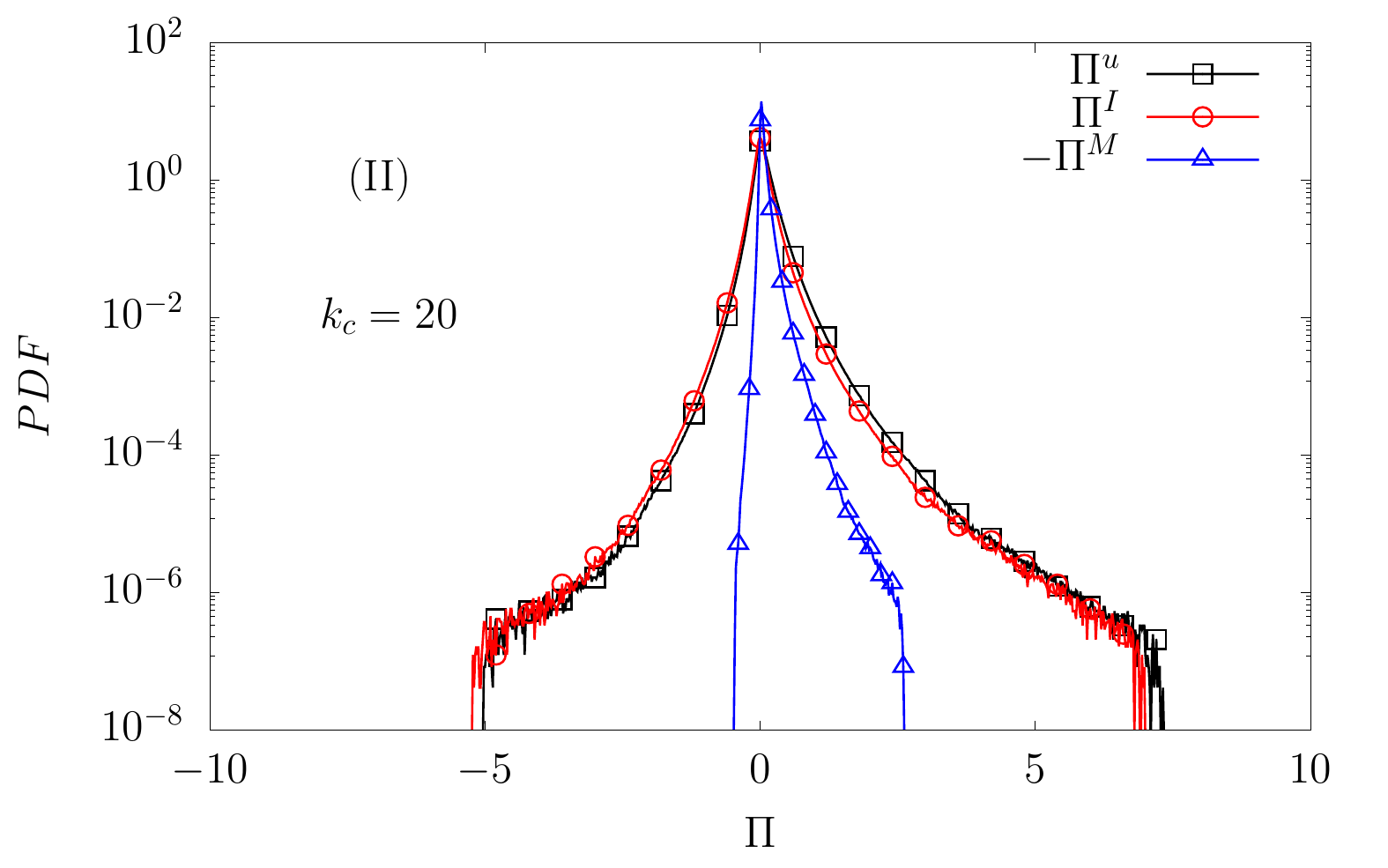}
  \includegraphics[width=0.5\textwidth]{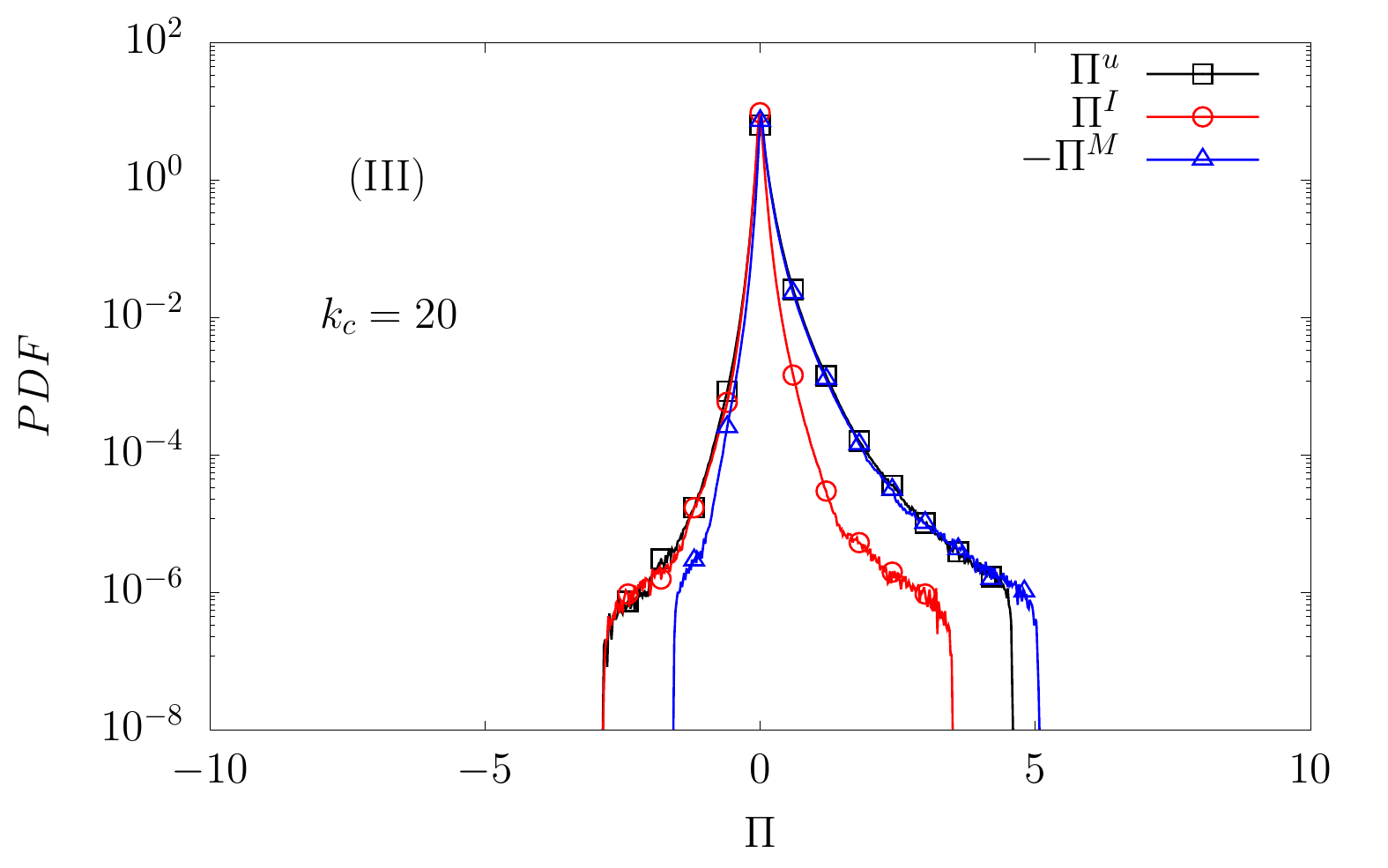}
  \caption{
        {Fluctuations of}
        $\Pi^{I}$ and $\Pi^{M}$ during
        kinematic (I), non-linear (II) and stationary (III) stages at $k_c=20$.
        }
  \label{fig:VELMAG-comp}
\end{figure}

{As shown in Fig.~\ref{fig:VELMAG_mean}(a), the mean inertial interscale energy
transfer is weakened in presence of a saturated dynamo. This partly occurs
through cancellations of forwards and inverse transfers since backscatter
events in $\Pi^{I}$ now occur more frequently as already discussed.
Additionally, an overall depletion of the fluctuations of $\Pi^{I}$ occurs, as
can be seen from  the comparison of the pdfs of $\Pi^{I}$ and $\Pi^{M}$ and
$\Pi^{u}$ presented in Fig.~\ref{fig:VELMAG-comp}.}

\begin{figure*}[htbp]
 \centering
 \includegraphics[width=0.9\textwidth]{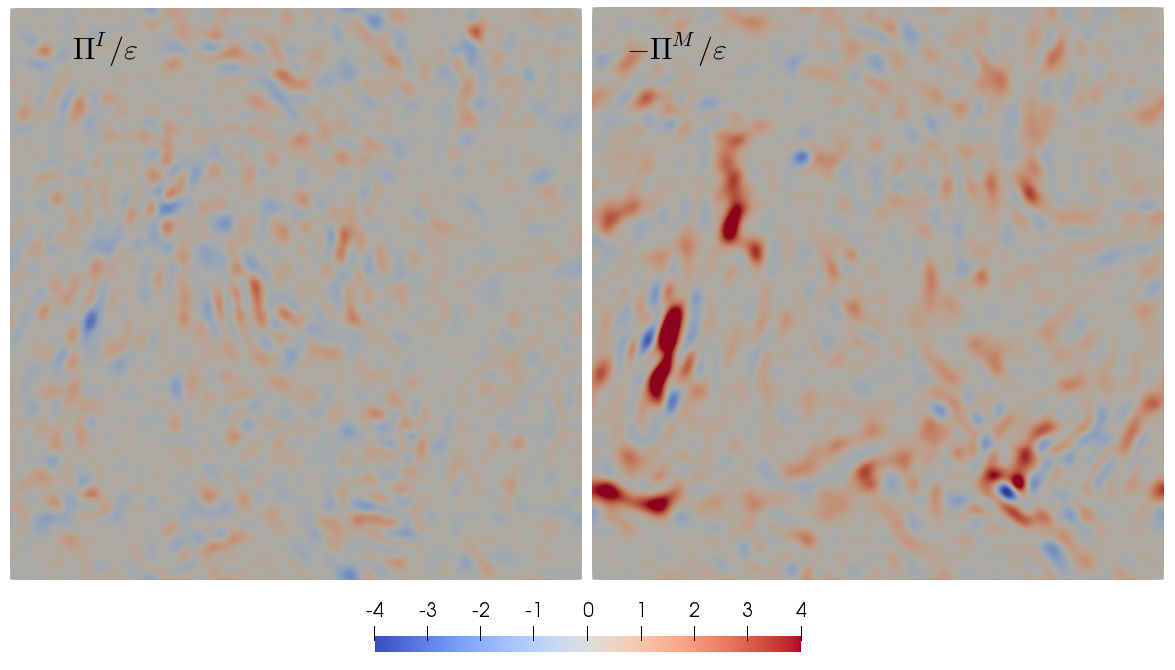}
 \includegraphics[width=0.9\textwidth]{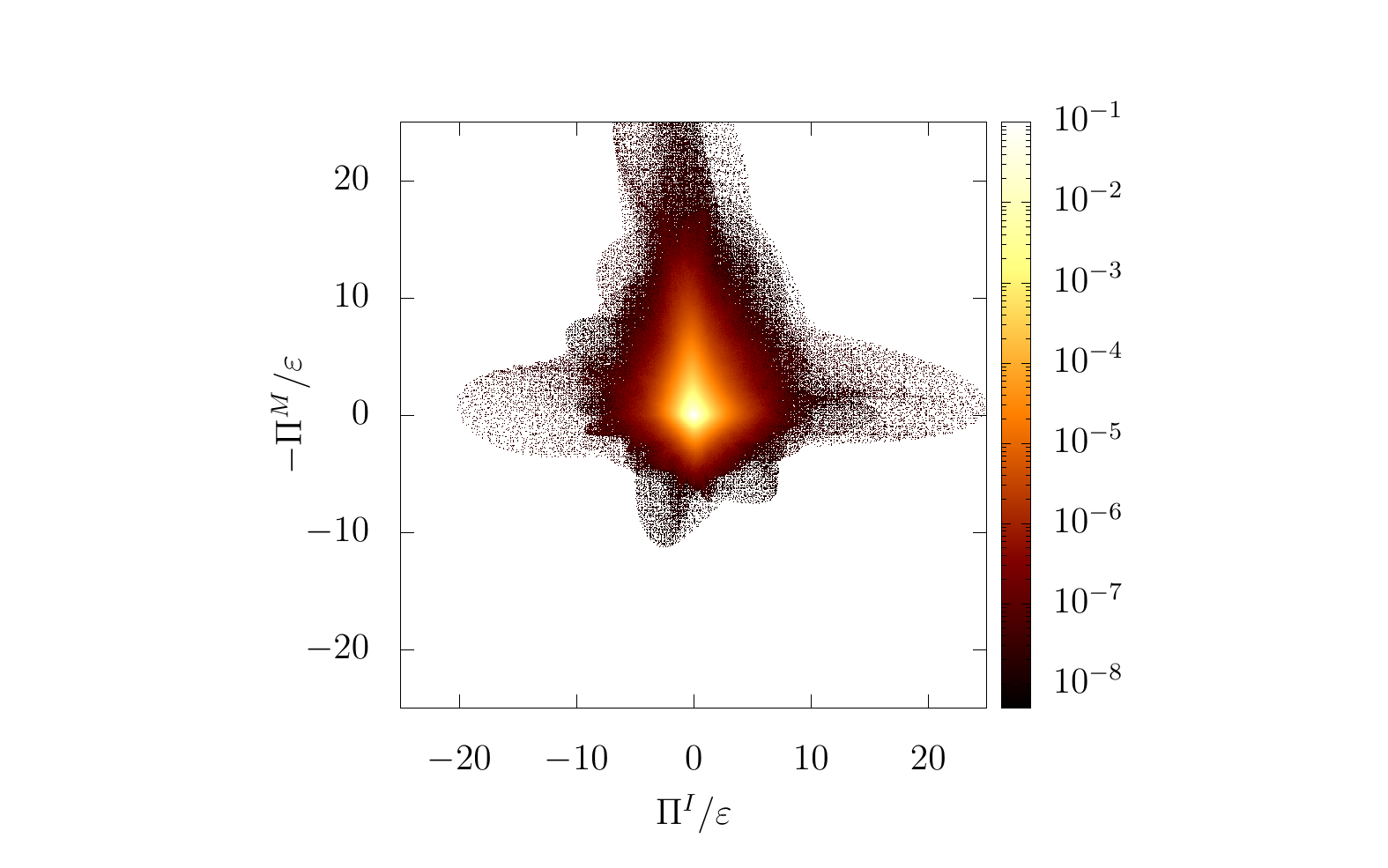}
 \caption{(Colour online)
   Top: Two-dimensional visualisations of the inertial and Maxwell SGS energy 
   transfers $\Pi^{I}$ (left) and $-\Pi^{M}$ (right)  at $k_c=20$ 
   in stage (III). 
   Positive values correspond to forward energy transfer while negative values
   incicate backscatter.
   {Bottom: Corresponding joint pdf of $\Pi^{I}$ and $-\Pi^{M}$.}
 }
 \label{fig:Visu_PiVM}
\end{figure*}

The clear forward transfer of energy {in stage (III)} associated with the Maxwell stress is also
visible in the 2D visualisations of $\Pi^{I}$ and $\Pi^{M}$ presented 
in {the top panels of} Fig.~\ref{fig:Visu_PiVM}. Unlike $\Pi^{I}$, $\Pi^{M}$ shows very intense and 
localized regions of forward transfer. 
{As discussed earlier, the pdf of $\Pi^{I}$ becomes quite symmetric in stage (III), indicating that
positive and negative fluctuations of $\Pi^{I}$ occur with similar probabilities. 
This is also visible in the visualisations, where we see regions of forward and inverse transfer which are of comparable intensity. The 
fluctuations of $\Pi^{I}$ also appear to be  much weaker than those of $\Pi^{M}$. 
Finally, we find that $\Pi^{I}$ and $\Pi^{M}$ \blue{have a relation of weak inverse proportionality} 
as can be seen from their joint 
pdf presented in the bottom panel of Fig.~\ref{fig:Visu_PiVM}. 
The latter suggests that the transfer of kinetic energy between resolved scales and SGS is more 
likely to occur {\em separately} through $\Pi^{I}$ or $\Pi^{M}$ rather than simultaneously through 
both.}  

\section{Conclusions}
\label{sec:conclusions}
In this paper, we investigated the different components of the SGS energy transfer 
through three stages of dynamo evolution considering mean and fluctuating properties. 
We decomposed the total SGS energy transfer in the components corresponding to
either the momentum or the induction equation, thus separating kinetic from 
magnetic SGS energy transfer. The kinetic SGS energy transfer was then further
split into an inertial component and a component originating from the Lorentz
force. By also distinguishing between the actual SGS energy transfers and 
residual contributions from interactions amongst the resolved scales, we got 
clear measurements of the fluctuating individual SGS energy transfers.

Concerning the velocity field, important differences are present between the
statistical properties of the inertial SGS energy transfer in presence of a
saturated dynamo and in the nonconducting case.  First, the kinetic energy
cascade is depleted in the saturated dynamo regime{, see Figs.~\ref{fig:VELMAG_mean}(a) and 
\ref{fig:VELMAG-comp}}.  Second, we find that the
pdf of the inertial SGS energy transfer becomes more symmetric and less
Gaussian than in the non-conducting case with wider tails suggesting more
extreme events also in terms of backscatter{, see Figs.~\ref{fig:VELMAG}(a) and 
\ref{fig:VELMAG-comp}}.  Third, we found quantitative evidence 
that the flatness of the
inertial SGS energy transfer has a weaker scale dependence, which 
\ml{suggests} that the velocity field \ml{may be} 
less intermittent in presence of a saturated small-scale
dynamo than in the nonconducting case{, see Fig.~\ref{fig:VELMAG}(c)}.  
This latter case deserves a more quantitative investigation 
by increasing the statistics and by extension of the involved scales.
Concerning the magnetic field, we find
that the pdf of the magnetic energy transfer is pretty symmetric in both the
nonlinear and the saturated dynamo regimes{, see Fig.~\ref{fig:VTOTBTOT}(b)}.  
In contrast, the SGS energy
transfer originating from the Maxwell stress in the momentum equation is
clearly skewed towards positive values{, see Figs.~\ref{fig:VELMAG}(b) and 
\ref{fig:VELMAG-comp}}.

\ml{In terms of fundamental results on interscale energy transfer in MHD turbulence, 
the filtering technique is a useful alternative to spectral approaches.  
According to analyses of shell-to-shell transfers, magnetic and velocity-field modes couple 
at disparate wave number shells \cite{Verma04,Alexakis05a,Mininni05a,Alexakis07}, 
leading to nonlocal contributions to the conversion of kinetic to magnetic energy in Fourier space. 
\blue{
As can be seen from Eqs.~\eqref{eq:evol_Eu} and \eqref{eq:evol_Eb}, 
the conversion of resolved-scale kinetic to
magnetic energy involves resolved scales only. Although not assessed
here, the energy conversion term for the SGS energies is
also closed in terms of the SGS \cite{Aluie17}. That is, the conversion terms do
not couple the resolved scales with the SGS. In summary, the
}	
filtering technique shows that energy conversion across the filter scale does 
not occur \cite{Aluie17}. 
The degree of locality of energy cascades is certainly affected by the presence of large-scale fields, such as in 
rotating turbulence, two-dimensional flows or in the presence of magnetic and kinetic helicity \cite{Alexakis18}, 
requiring further analysis of the effect of SGS closures on higher-order statistics \cite{Linkmann18}. 
Hence, separate {\em a-priori} studies are required in order to provide guidance for LES modeling in such cases,
as e.g for large-scale dynamos \cite{Kessar16}.  
}

In terms of guidance for LES modelling, the symmetry  of the
magnetic SGS energy transfer pdf implies that backscatter events are important,
which calls applications of dissipative models for the stresses in the
induction equation into question. For the momentum equation, a similar situation
occurs for the inertial SGS energy transfer in the saturated stage of the dynamo.
As a result, while a dissipative model for the inertial stresses may be suitable 
during the kinematic stage, a more sophisticated approach is required to adequately 
capture the increased backscatter in the nonlinear and saturated stages.   
On the other hand, dissipative models would be well suited for the Maxwell stress in both 
nonlinear and saturated stages.  
{Finally, we find that the \blue{correlation between the}
individual SGS energy transfers appears to be \blue{of inverse proportionality} 
in the saturated stage. 
This holds for $\Pi^u$ and $\Pi^b$ and also for $\Pi^I$ and $\Pi^M$. That is, the energy transfers in the different channels 
appear to occur separately, which should be taken into account in the design of more sophisticated LES models for MHD.}
However, measurements of the correlations between the different SGS energy transfers {at higher Reynolds numbers} 
need to be carried out in order to better quantify the effect.


\section*{Acknowledgements}
The research leading to these results has received funding from the European Union's Seventh
Framework Programme (FP7/2007-2013) under grant agreement No. 339032.

\appendix

\section{Comparison between the SGS and Leonard energy transfers}
\label{app:Leonard}
\begin{figure}[htbp]
  \centering
  \includegraphics[width=0.5\textwidth]{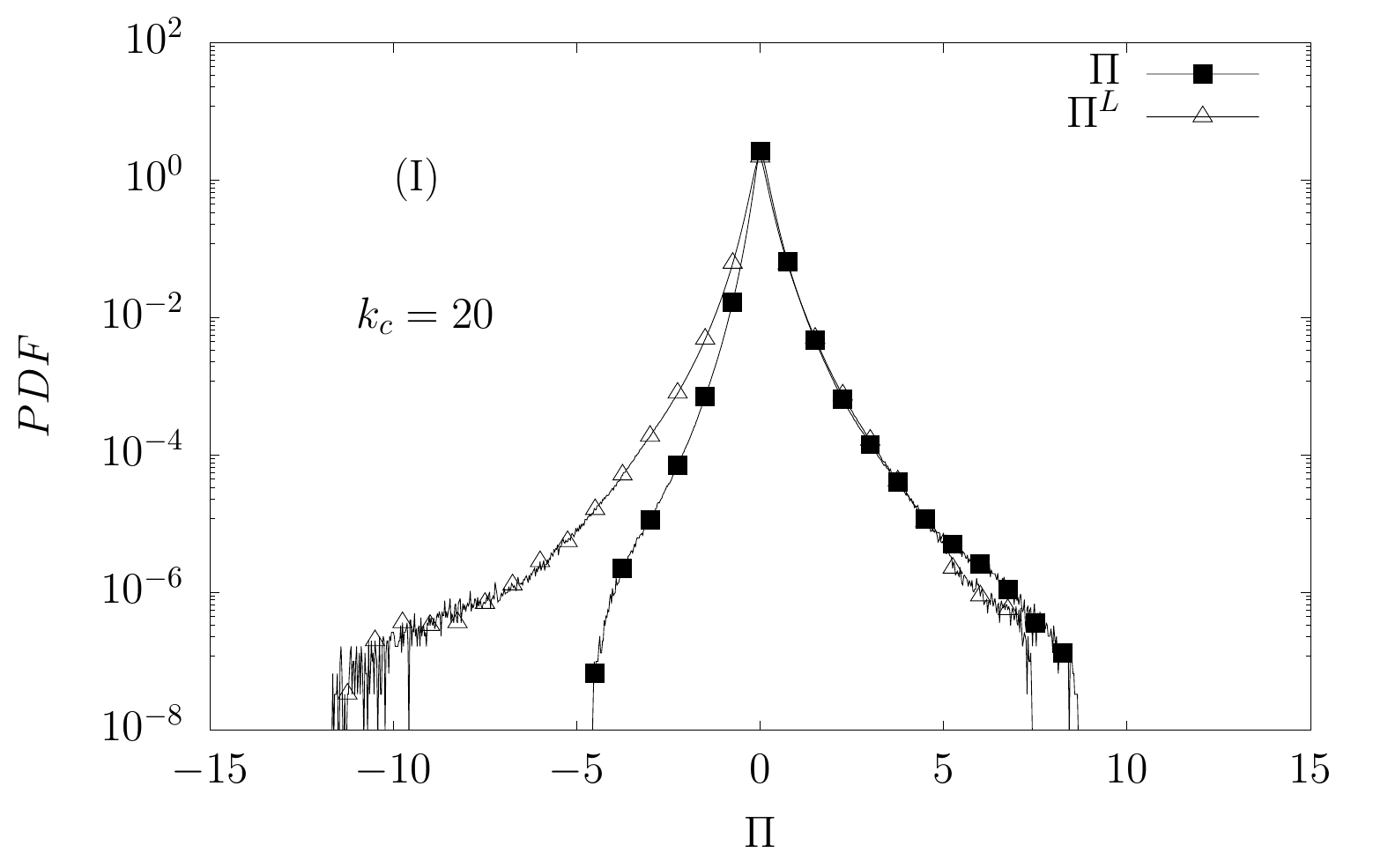}
  \includegraphics[width=0.5\textwidth]{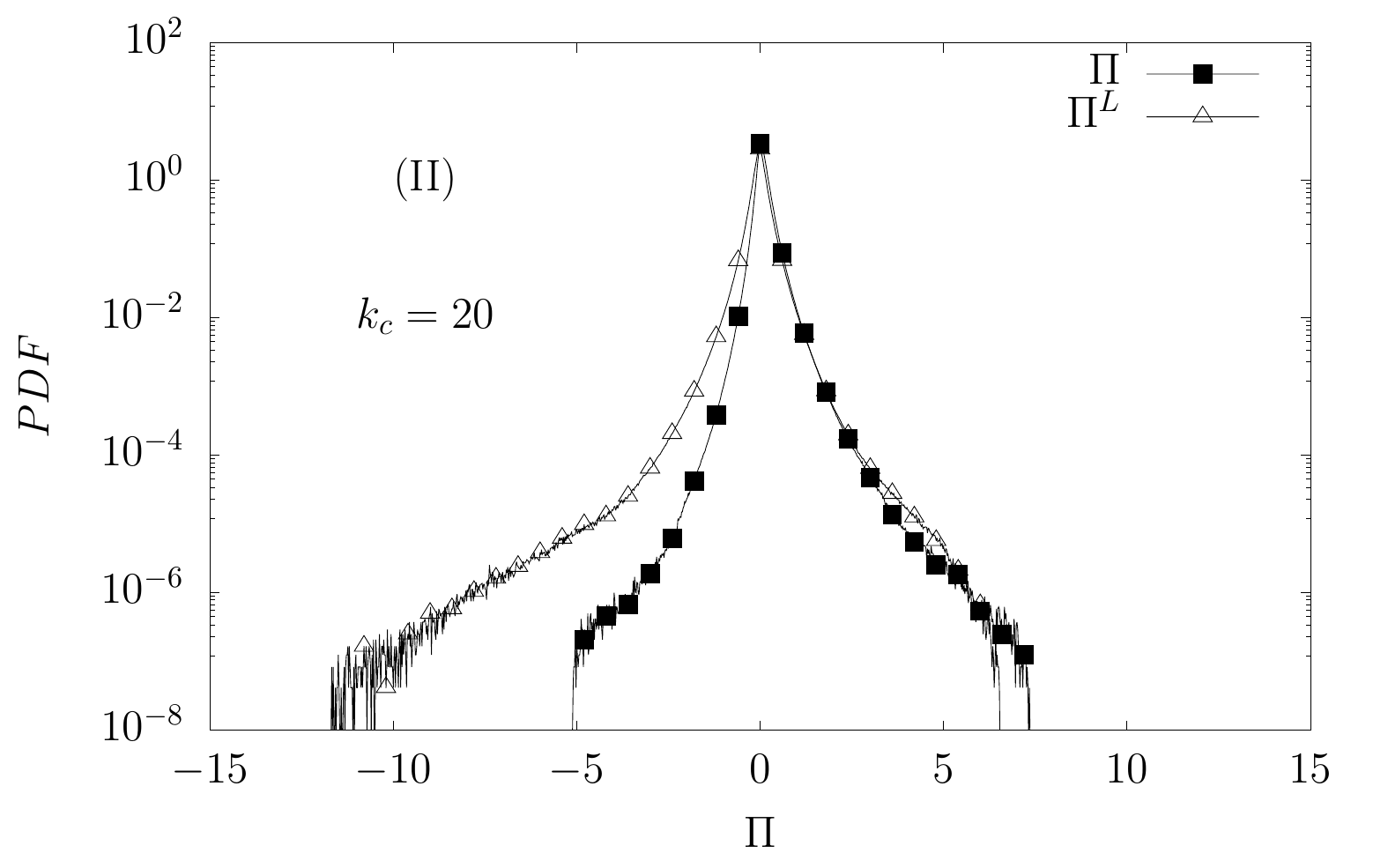}
  \includegraphics[width=0.5\textwidth]{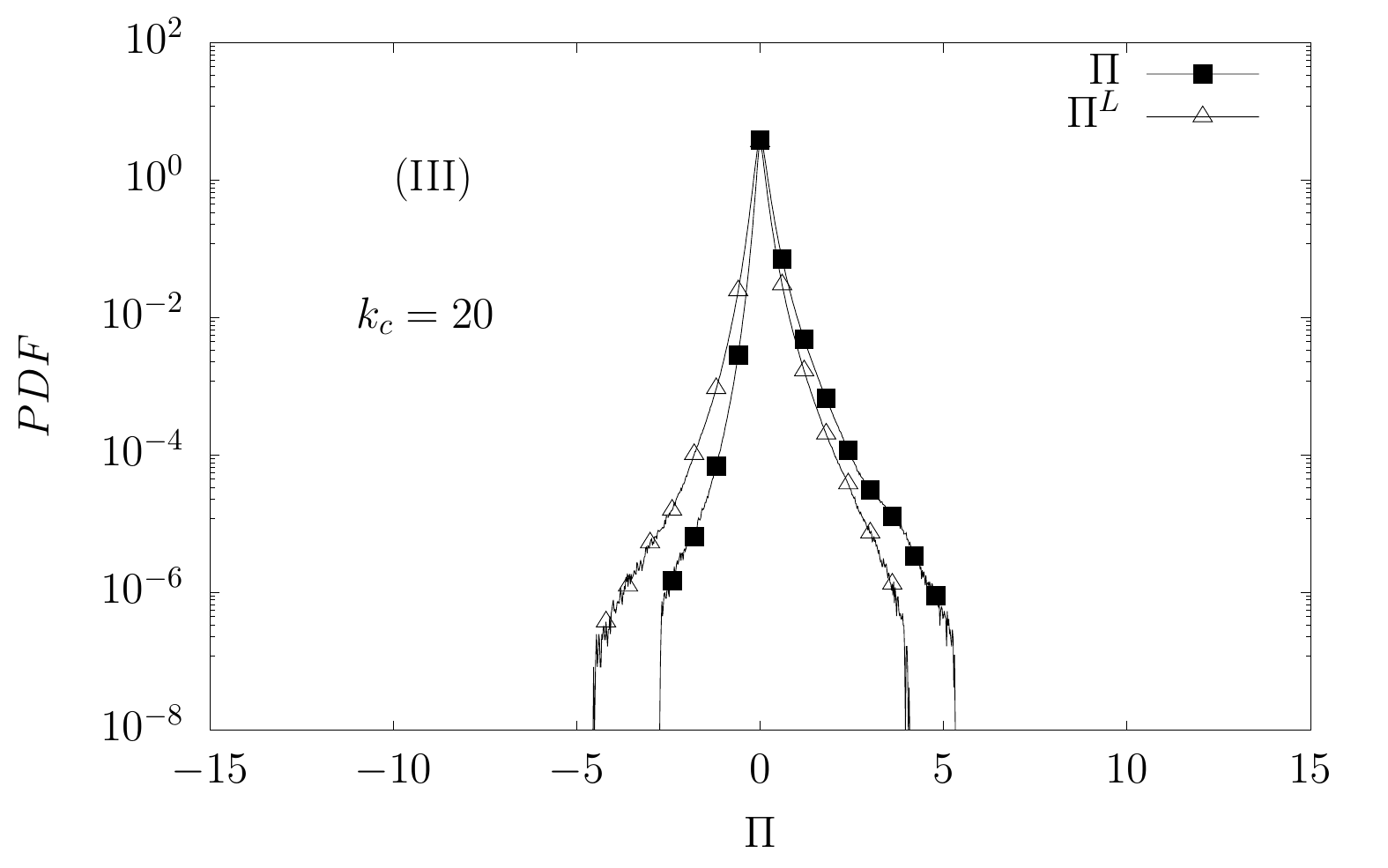}
  \caption{pdfs of $\Pi$ and $\Pi^L$ at the kinematic stage (I), non-linear stage (II) and the stationary stage (III), with $k_c=20$.}
  \label{fig:TOT_LEO}
\end{figure}

\begin{figure*}[htbp]
  \centering
  \includegraphics[width=0.48\textwidth]{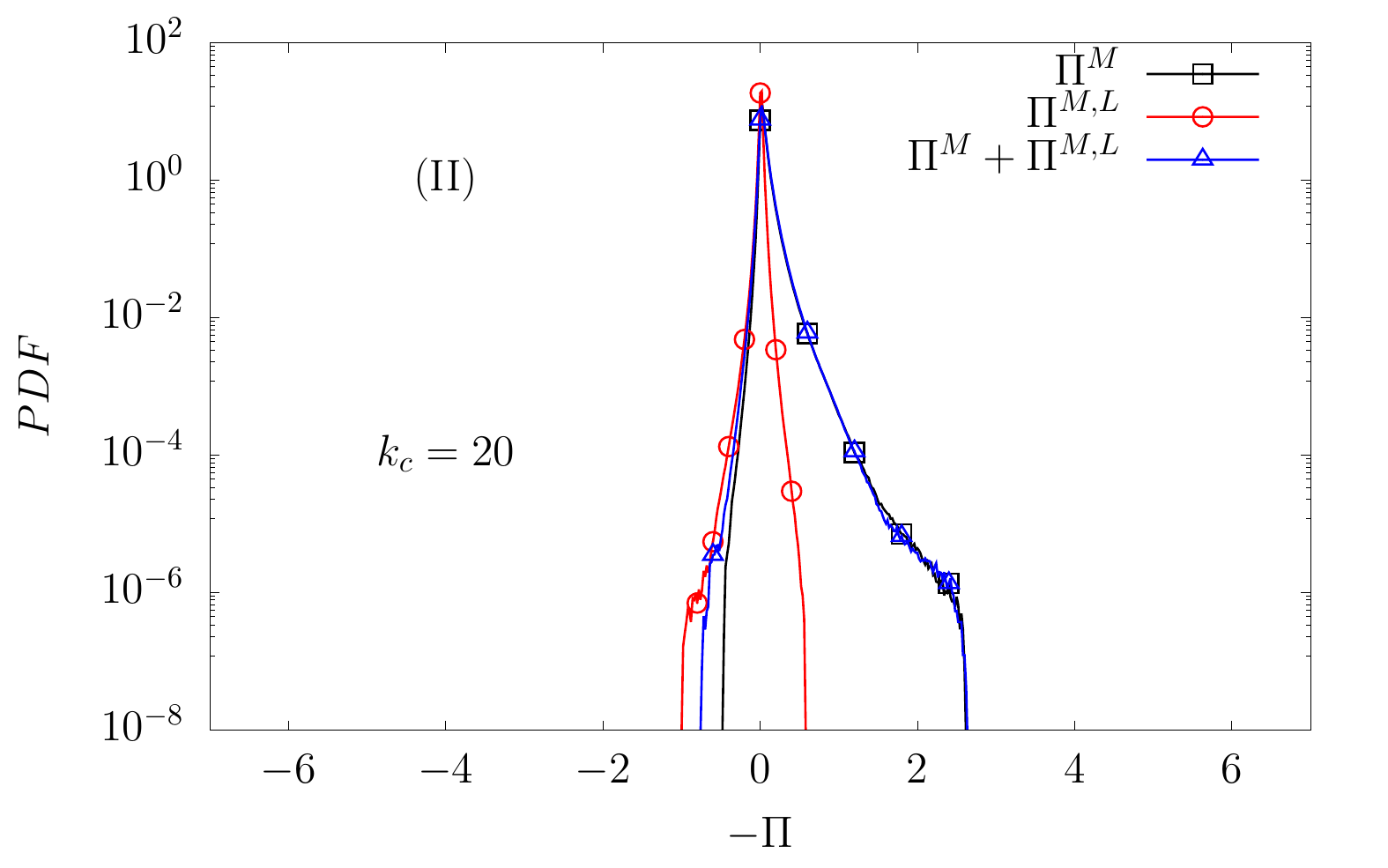}
  \includegraphics[width=0.48\textwidth]{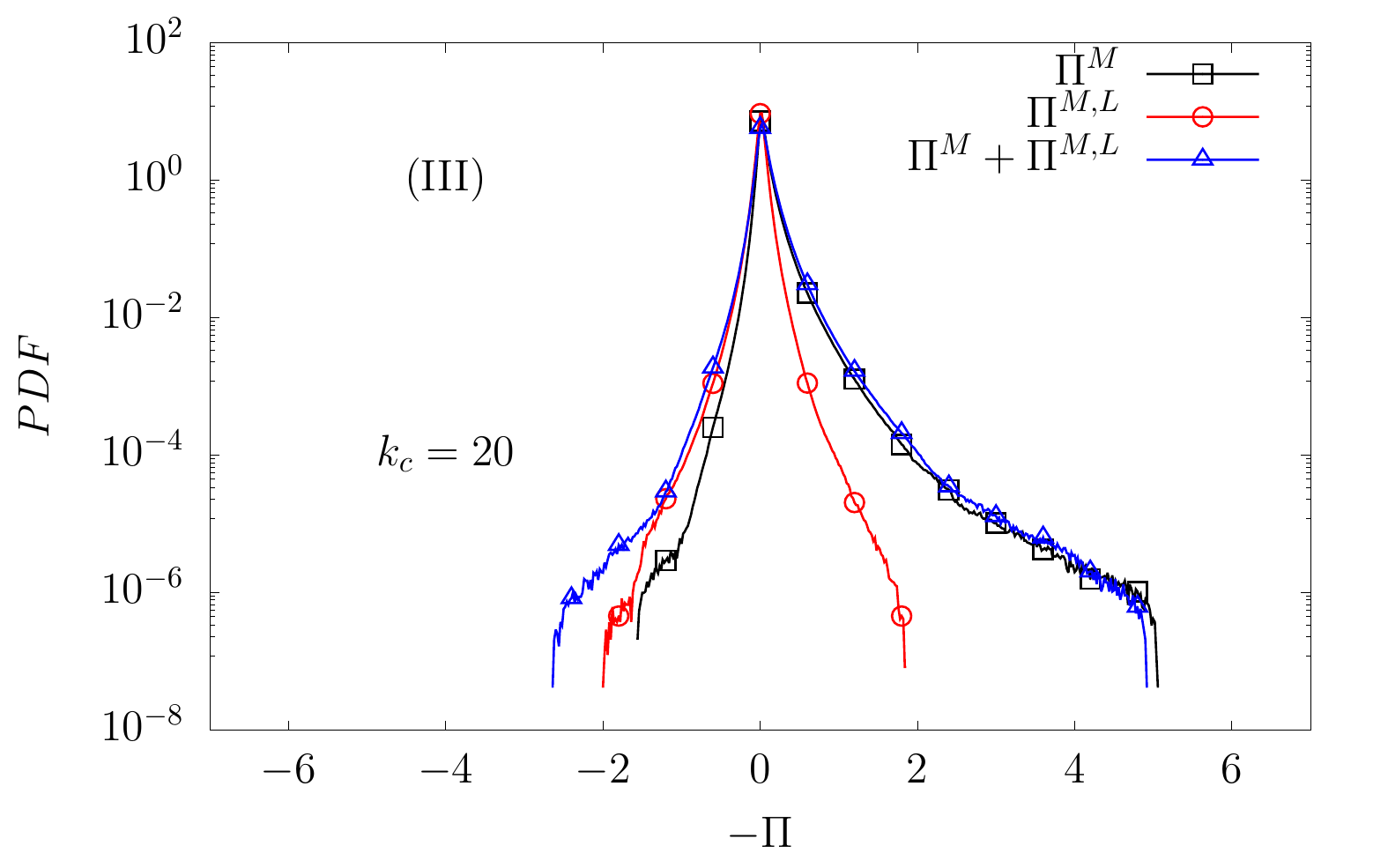}
  \caption{pdfs of the  Maxwell SGS energy transfers $\Pi^{M}$ and $\Pi^{M,L}$ and their sum in the 
           nonlinear (II) and the stationary (III) stages, with $k_c=20$.}
\label{fig:MAG_LEO}
\end{figure*}

\begin{figure*}[htbp]
  \centering
  \includegraphics[width=0.48\textwidth]{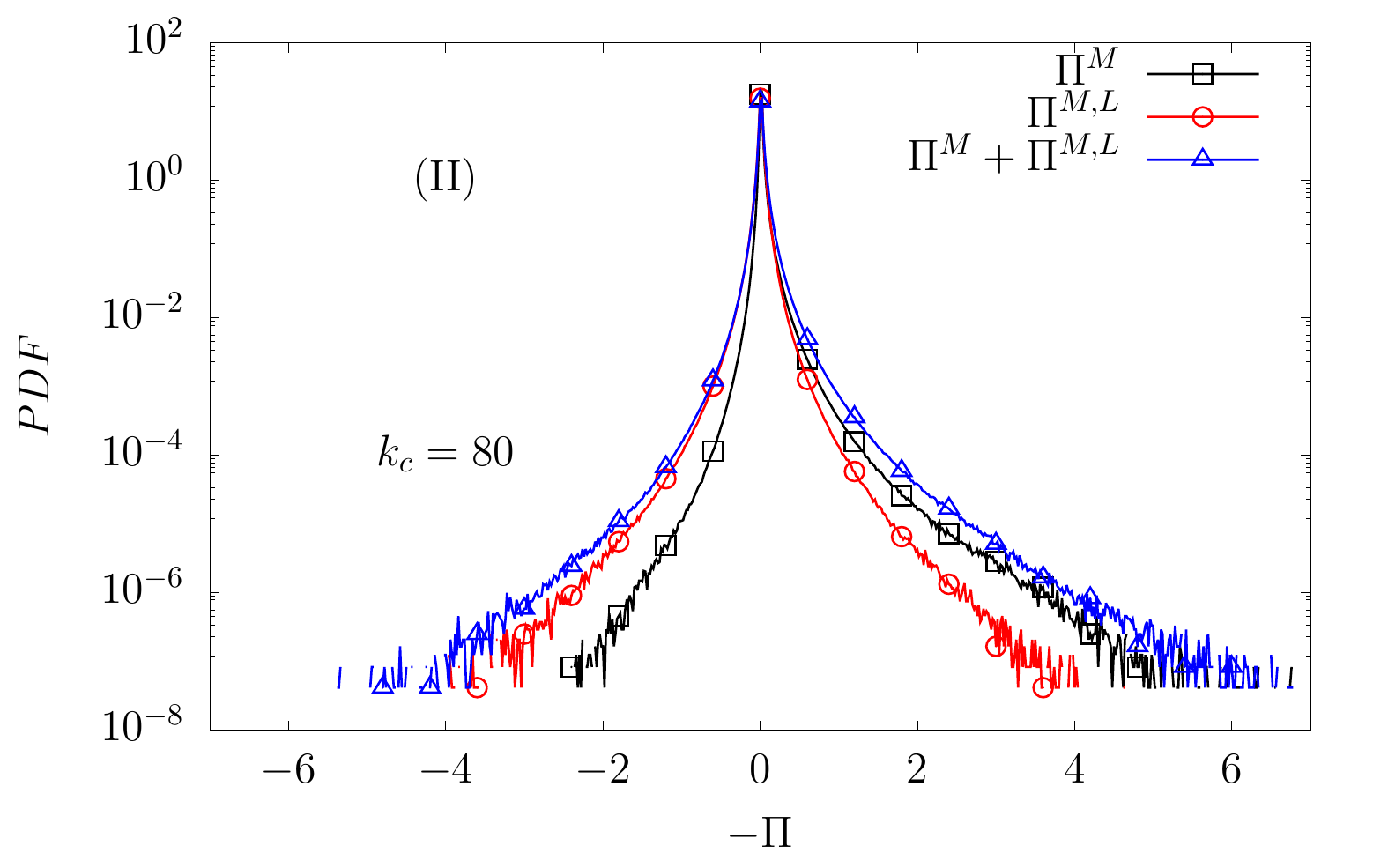}
  \includegraphics[width=0.48\textwidth]{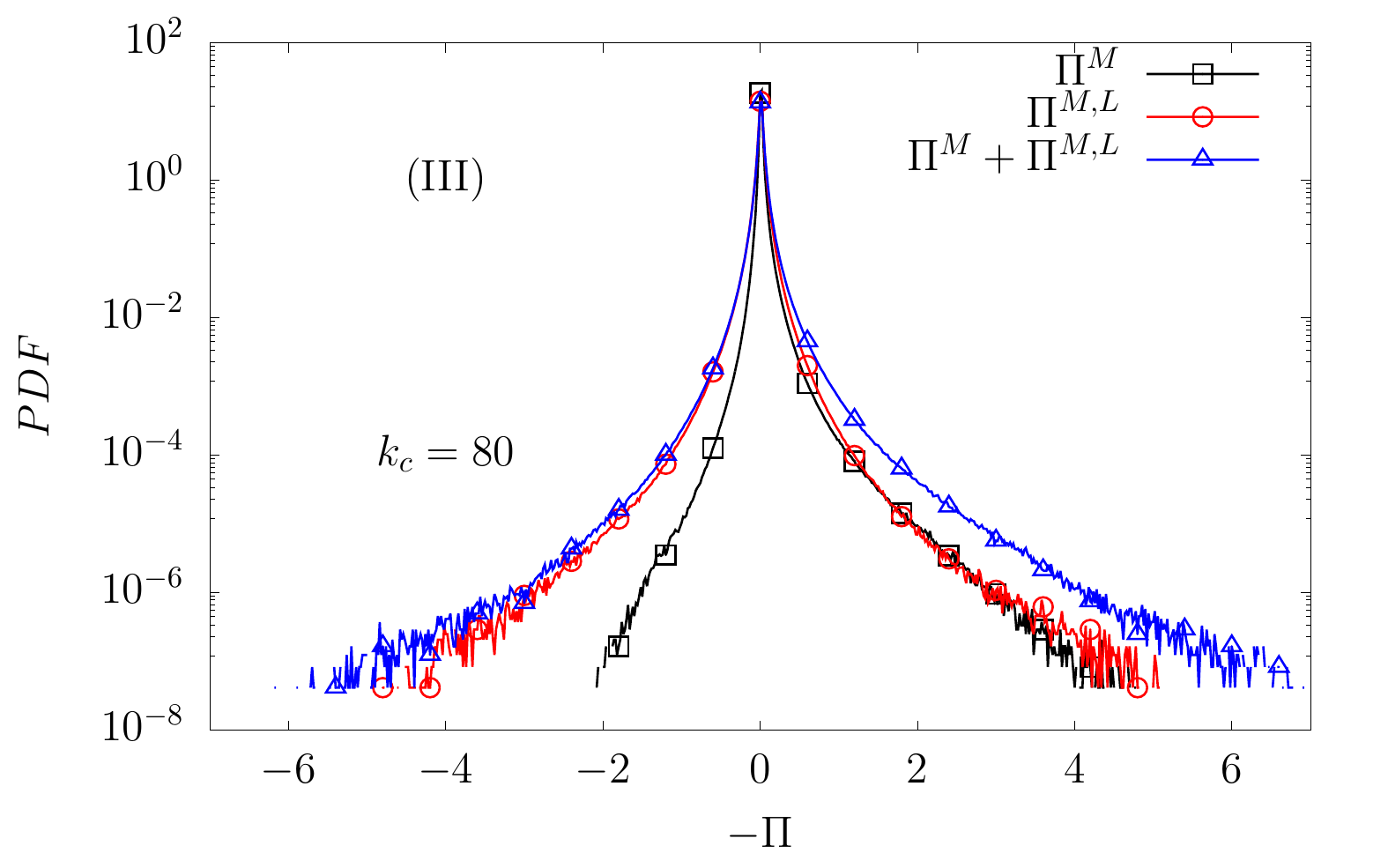}
  \caption{pdfs of the  Maxwell SGS energy transfers $\Pi^{M}$ and $\Pi^{M,L}$ and their sum in the            nonlinear (II) and the stationary (III) stages, with $k_c=80$.}
  \label{fig:MAG_LEO_kc80}
\end{figure*}

The Leonard components of the individual energy transfer terms were defined in 
Eqs.~\eqref{eq:Leo_u}-\eqref{eq:Leo_b}. As mentioned before, the Leonard transfers
do not contribute to the SGS energy transfer as they are closed in terms of 
the resolved fields. Furthermore, it can be shown that they vanish under spatial
averaging. The latter suggests that forward and backward energy transfer should be 
more or less equally likely. Figure \ref{fig:TOT_LEO} presents comparisons between 
the actual SGS energy transfer $\Pi$ and its Leonard component  $\Pi^L$
at $k_c = 20$ and during stages (I)-(III). As can be seen, $\Pi^L$
is indeed more symmetric than $\Pi$ in all cases. This situation is also present
for the Maxwell energy transfers $\Pi^{M}$ and $\Pi^{M,L}$ shown in Fig.~\ref{fig:MAG_LEO}
for the nonlinear and stationary stages of dynamo evolution. At least in stage (III), a measurement 
of $\Pi^{M} + \Pi^{M,L}$ instead of $\Pi^{M}$ would have resulted in a more pronounced left tail of the pdf, leading to the consclusion of more backscatter being present in the Maxwell SGS 
transfer than there actually is. {In Fig.~\ref{fig:MAG_LEO_kc80} the same measurements of the Maxwell energy transfers are presented at a different cutoff closer to the dissipation range, namely $k_c=80$. At this scale, all pdfs show a higher probability to measure extreme events of energy transfer. However, the pdf of the Maxwell energy transfer $\Pi^M$ remains clearly skewed towards the right, 
which suggests that the extreme events remain correlated to the direction of the mean energy flux, even though they become more than two orders of magnitude larger compared to the mean value. 
Moreover, as already observed in Fig.\ref{fig:TOT_LEO} from the pdfs of the total energy transfer, this information is not accessible
through a measurement of 
the sum between $\Pi^M$ and $\Pi^{M,L}$, because 
the Leonard term is completely symmetric and large enough to dominate the left tail of the PDF. 
The same results are valid in both the nonlinear and the stationary stage.}

\bibliographystyle{unsrt}
\bibliography{refs,ref2,ref3}

\end{document}